\documentclass[reprint, nofootinbib,showpacs,superscriptaddress,11pt]{revtex4}

\usepackage{blindtext}
\usepackage{hyperref}

\usepackage{scrextend}

\usepackage[makeroom]{cancel}
\usepackage{footnote}
\usepackage{slashed}
\usepackage{lipsum}
\usepackage{hyperref}
\usepackage{amssymb,amsmath}
\usepackage[dvips,final]{graphicx}
\usepackage{url}
\usepackage{subfigure}
\usepackage{textcomp}
\usepackage{graphicx}
\usepackage{bm}
\usepackage{dcolumn}
\usepackage[usenames]{color}
\usepackage{scrextend}
\usepackage{amssymb}
\usepackage{amsmath}
\usepackage{bm}
\usepackage{amsthm}
\usepackage{amsopn}

\def \k{\mathbf{ k}}
\def \p{\mathbf{ p}}
\def \q{\mathbf{ q}}
\def \e{\mathbf{ e}}
\def \l{\left}
\newcommand{\abs}[1]{\left| #1\right|}
\def \r{\right}
\def \x{\mathbf{ x}}
\def \K{\mathbf{ K}}
\def \S {\mathcal{S}}

\def \tmes{t_{\textrm{mes}}}
\def \tmic{t_{\textrm{mic}}}
\def  \g{(\textrm{g})}

\def\comment#1{}
\newcommand{\Tr}{\,\mathrm{Tr}}

\def\beq{\begin{equation}}
\def\eeq{\end{equation}}
\def\bea{\begin{eqnarray}}
\def\eea{\end{eqnarray}}

\makeatletter
\usepackage{ulem}
\usepackage{cancel}
\usepackage{color}

\makeatother

\begin{document}

\title{ \Large Non-Markovian open quantum system approach to the early universe: I. Damping of gravitational waves by matter }

\author{M. Zarei}
\email[]{m.zarei@iut.ac.ir}
\affiliation{Department of Physics, Isfahan University of Technology, Isfahan 84156-83111, Iran}
\affiliation{ICRANet-Isfahan, Isfahan University of Technology, 84156-83111, Iran}
\affiliation{Dipartimento di Fisica e Astronomia “Galileo Galilei” Universita` di Padova, 35131 Padova, Italy}
\affiliation{INFN, Sezione di Padova, 35131 Padova, Italy}

\author{N. Bartolo}
\email[]{nicola.bartolo@pd.infn.it}
\affiliation{Dipartimento di Fisica e Astronomia “Galileo Galilei” Universita` di Padova, 35131 Padova, Italy}
\affiliation{INFN, Sezione di Padova, 35131 Padova, Italy}
\affiliation{INAF - Osservatorio Astronomico di Padova, I-35122 Padova, Italy}

\author{D. Bertacca}
\email[]{daniele.bertacca@gmail.com}
\affiliation{Dipartimento di Fisica e Astronomia “Galileo Galilei” Universita` di Padova, 35131 Padova, Italy}

\author{S. Matarrese}
\email[]{sabino.matarrese@pd.infn.it}
\affiliation{Dipartimento di Fisica e Astronomia “Galileo Galilei” Universita` di Padova, 35131 Padova, Italy}
\affiliation{INFN, Sezione di Padova, 35131 Padova, Italy}
\affiliation{INAF - Osservatorio Astronomico di Padova, I-35122 Padova, Italy}
\affiliation{Gran Sasso Science Institute, I-67100 L'Aquila, Italy}

\author{A. Ricciardone}
\email[]{angelo.ricciardone@pd.infn.it}
\affiliation{Dipartimento di Fisica e Astronomia “Galileo Galilei” Universita` di Padova, 35131 Padova, Italy}
\affiliation{INFN, Sezione di Padova, 35131 Padova, Italy}

\date{\today}

\begin{abstract}

By revising the application of the open quantum system approach to the early universe and extending it to the conditions beyond the Markovian approximation, we obtain a new non-Markovian quantum Boltzmann equation. Throughout the paper, we also develop an extension of the quantum Boltzmann equation to describe the processes that are irreversible at the macroscopic level. 
This new kinetic equation is, in principle, applicable to a wide variety of processes in the early universe. For instance, using this equation one can accurately study the microscopic influence of a cosmic environment on a system of cosmic background photons or stochastic gravitational waves. 
In this paper, we apply the non-Markovian quantum Boltzmann equation to study the damping of gravitational waves propagating in a medium consisting of decoupled ultra-relativistic neutrinos. 
For such a system, we study the time evolution of the intensity and the polarization of the gravitational waves. It is shown that, in contrast to intensity and linear polarization that are damped, the circular polarization (V-mode) of the gravitational wave (if  present) is amplified by propagating through such a medium.

\end{abstract}

\maketitle

\section{Introduction}

Master equations are a powerful tool for studying the dynamics of density matrices associated with an Open Quantum System (OQS) \cite{Breuer,Lidar}. They allow for describing the relevant degrees of freedom of the system, which evolve under the influence of all other degrees of freedom collectively called the environment. An incredibly simple situation occurs when a time-local master equation can describe the system with constant dissipation rates. This results in Markovian evolution, where the knowledge of the density matrix at a given time is sufficient to predict all future observables meaning that the environment has no memory. This type of master equation is referred to as the Markovian equation. Compared to the whole problem of describing all the degrees of freedom of system and environment together, a Markovian master equation that controls only the system's degrees of freedom is an immense simplification. Such a drastic reduction in complexity is usually costly. In this case, the price comes in terms of strong approximations, which are not always justified. 
The study of such approximations is thus of great importance, and in fact, there is a large body of literature that addresses these issues (see \cite {Breuer,Lidar} for review). Many recent studies have focused on revisiting the question of the validity of the widely used Markovian quantum master equations and have developed techniques to investigate non-Markovian dynamics of open systems \cite{Cattaneo,Vega,Breuer-paper,Rivas,Hofer}. 

The formalism of the Quantum Boltzmann Equation (QBE) was initially developed to study the time evolution of the intensity and the polarization of the Cosmic Microwave Background (CMB) photons \cite{Kosowsky:1994cy,Bavarsad:2009hm,Bartolo:2018igk,Bartolo:2019eac,Bartolo:2020htk}. One way to formulate such a quantum kinetic equation is to use the open quantum system approach. Starting from a master equation describing the CMB photon density matrix dynamics, applying the Born-Markov approximation, and finally, after taking the operator expectation values, the QBE arises \cite{Kosowsky:1994cy,Bavarsad:2009hm,Bartolo:2018igk,Bartolo:2019eac,Bartolo:2020htk,Fidler:2017pkg}. 
The QBE is an accurate tool to extract the effects of microscopic interactions on the macroscopic properties of the intensity and the polarization of the CMB radiation. 
Since the photon and graviton density matrices have the same number of degrees of freedom, the same formalism of the QBE can be applied to study the propagation of the Stochastic Gravitational-Wave Background (SGWB) in a medium.

The SGWB represents a laboratory where the QBE can be applied. In the near future, we expect that Gravitational-Wave (GW) interferometers, like Advanced LIGO/Virgo/KAGRA, LISA or Einstein Telescope~\cite{Audley:2017drz, Kawamura:2006up, Evans:2016mbw, Sathyaprakash:2011bh, Maggiore:2019uih}, will be sensitive enough to measure the astrophysical background produced by many unresolved GW sources, like Black Holes and Neutron Stars, and the cosmological background coming from early universe sources, such as inflation, phase transitions, topological defects, etc. (see e.g. \cite{Regimbau:2011rp, Guzzetti:2016mkm, Caprini:2018mtu} for reviews).
 SGWBs are then powerful tools to shed light on astrophysics, cosmology and fundamental physics~\cite{Maggiore:1999vm,  Bartolo:2016ami, Maggiore:2019uih,Barausse:2020rsu}. 
 Among all these effects, we will investigate the GW damping due to propagation in a dissipative environment.
 
 The seminal work of Hawking \cite{Hawking:1966qi} has revealed that GWs do not interact with a perfect fluid in the absence of dissipative processes. However, in a viscous medium, the energy of GWs is converted into heat, without provoking macroscopic motions of the medium \cite{Hawking:1966qi,Gayer:1979ff,Polnarev,Szekeres:1971ss,Weinberg:1972kfs}. 
A medium with a dynamical viscosity coefficient $\eta$ could absorb the GW at a rate of absorption \cite{Hawking:1966qi,Weinberg:1972kfs},
\beq 
\Gamma_{GW}=16\pi G \eta~. 
\eeq 
This result is valid only if the collision frequency in the matter \footnote{The collision frequency indicates the number of collisions per unit time. This collision frequency is defined as the inverse of the (mean) time $\tau$ between two collisions, known as the scattering time or relaxation time.}
is much greater than the frequency of the GW. 
A well-known effect given by decoupled relativistic neutrinos on the CMB angular power-spectrum is the damping due to their anisotropic stress of the amplitude of the GW spectrum by~$35\%$~\cite{Weinberg:2003ur} (see also \cite{Pritchard:2004qp,Baym:2017xvh,Mangilli:2008bw,Watanabe:2006qe,Dicus:2005rh,DallArmi:2020dar,Stefanek:2012hj,Miron-Granese:2020hyq}). Such a damping becomes quite large in the frequency region between $10^{-16}$ Hz and   $10^{-10}$ Hz~\cite{Weinberg:2003ur,Pritchard:2004qp,Baym:2017xvh,Mangilli:2008bw,Watanabe:2006qe,Dicus:2005rh,DallArmi:2020dar,Stefanek:2012hj}. In a similar way decoupled relativistic particles in the early universe affect the anisotropies of the cosmological background of GWs becoming testable predictions for future GW interferometers~\cite{DallArmi:2020dar}. 
The opposite case, in which the collision frequency is very small (so that one can consider the matter as collisionless
matter), has been studied in \cite{Flauger:2017ged}, resulting in no observable effect except perhaps for cosmological sources. 
The quantum effects, such as the absorption and stimulated emission of low-frequency gravitational waves by a hot ionized gas have also been studied in \cite{Flauger:2019cam}.
As shown in this paper, such effects are well captured by the QBE approach, which can then be used to extract predictions for GW experiments.
 
We will apply the approach of the QBE to study the interaction of the SGWB with a thermal ultra-relativistic fermion bath. We also rederive the GW damping effect for the case that fermions are decoupled relativistic neutrinos \cite{Weinberg:2003ur}. 
The standard form of QBE is based upon an open quantum system. In this case, the gravitational wave can be viewed as an open system ($\mathcal{S}$) that interacts with the environment 
$\mathcal{E}$ (here, the fermion bath). The time evolution of $\mathcal{S}$ is obtained from the total $\mathcal{S}$+$\mathcal{E}$
dynamics by eliminating (i.e., integrating over) the $\mathcal{E}$ degrees of freedom. 
It may be possible to safely ignore the details of the bath dynamics
and use an effective description of this medium as a classical viscus background. 
It is usually assumed that the associated correlations are sufficiently small and the interaction takes place in such a way that the $\mathcal{S}$-$\mathcal{E}$ coupling is weak. This is known as a Born approximation. The back-reaction of $\mathcal{E}$ on $\mathcal{S}$ is also ignored. This assumption is known as the Markov approximation. To achieve the Markov approximation, we must ignore all references to history. The whole approximation is known as the Born-Markov approximation and the process is called a Markovian process. 
However, we will show that to rederive the GW damping effect, we cannot ignore the back-reaction effects of the ultra relativistic fermion, and hence, we must use a non-Markovian approach. 
In this work, we revise the interaction of the GWs with a dissipative environment of the decoupled ultra-relativistic fermions in the light of the concept of the OQS and the QBE. Moreover, to describe a damping phenomenon using the QBE, it is necessary to extend this equation to irreversible processes.

 Besides this example, there are many other applications that could
benefit from this appraoch.
The basic question of how the macroscopic irreversible behavior of a system arises
from the microscopic dynamics of quantum fields is a fundamental question,
with several diverse applications in the early universe, heavy-ion collisions, fluid dynamics, and condensed matter physics \cite{Danielewicz:1982kk,Bruus,Rammer,Kamenevej,Berges:2004yj}. 
It's been a  long time since physicists became interested in analyzing the properties of fluids employing a quantum field theoretical approach \cite{Jeon:1995zm}.
One of the important parameters in this framework is the time-scale between collisions in a viscous fluid, known as the {\it scattering} or {\it relaxation time}.
Extensive work has been carried out to develop fundamental methods for calculating the relaxation time \cite{Boyanovsky:1996xx,Calzetta:1999ps,Arnold:2000dr,Blaizot:2001nr,Policastro:2001yc, Manshouri:2020avm}.  One of the proposed approaches is known as {\it Schwinger-Keldysh} or {\it closed-time-path} formalism that is a useful tool to study the non-equilibrium initial value problems 
\cite{Schwinger:1960qe,Keldysh:1964ud}. 

By using the QBE we will be able to study the microscopic interactions of the system with the environment. Generally, either the system or the environment has an infinite number of degrees of freedom, and calculating the microscopic interaction can be difficult and impractical. Various techniques have been developed to investigate these microscopic effects. In the QBE approach, we use the techniques of quantum field theory, which solve these problems to a great extent. Furthermore, after taking the expectation value over the relevant operators, in the manner that will be described in the text, a macroscopic description of the system emerges.

The paper is organized as follows: In Section II, we provide a description and detailed comparison of the Markovian and non-Markovian approaches to the Quantum Boltzmann Equation, while in Section III, we deal with the implementation of the non-Markovian approach to the QBE to the study of the absorption of soft gravitons by a fermion bath. Section IV is devoted to the comparison of our method with the classical results by Weinberg \cite{Weinberg:2003ur} to this problem and the study of the evolution of the GW polarization in this process. Our main conclusions are given in Section V. The explicit calculation of expectation values of relevant quantities is presented in Appendix A.


\section{Markovian versus non-Markovian QBE}

In this section, we first review the Markovian and the non-Markovian master equations. Then, we will derive an extension of the QBE equation beyond the Markovian approximation. The non-Markovian QBE can be used to study various phenomena in the early universe. Here, we will utilize this equation to explain the damping of GWs. This section is devoted to the derivation of the QBE in three cases: (i) The process is reversible, and the back-reaction of the environment on the system is small (Markov approximation). (ii) Back-reaction is small, but the process is irreversible. (iii) Back-reaction impact cannot be ignored (non-Markov approximation), and the process is irreversible. 

\subsection{Master equation: Born-Markov and secular approximations}

In the OQS approach, the {\it open quantum system} $\mathcal{S}$ (for example, soft gravitons) is coupled to another quantum system (fermion bath in our case) called {\it the environment} $\mathcal{E}$. The free Hamiltonian of the system and the environment are described by $H_{\mathcal{S}}$ and $H_{\mathcal{E}}$, respectively. The interaction Hamiltonian $H_{\textrm{int}}$ describes the interaction between the system and the environment. The total microscopic Hamiltonian of such an open quantum system is given by 
\beq 
H_{\mathcal{S}+\mathcal{E}}=H_{\mathcal{S}}+H_{\mathcal{E}}+H_{\textrm{int}}~. 
\eeq 
$H_{\textrm{int}}$ is the interaction Hamiltonian, which causes dissipation or dephasing phenomena, where the former refers to both losses of energy and decoherence, and the latter refers to causing - at least in the uncoupled case - pure decoherence but no energy leak. The Markovian master equations can be derived in the weak-coupling limit of the system-environment interaction. Therefore, we can introduce a general dimensionless perturbation parameter $g$ that refers to the coupling constant, such that $H_{\textrm{int}}=\mathcal{O}(g)$. 
It is assumed that the system and the environment are coupled 
so weakly that the state of the environment is almost not perturbed by the 
coupling with the system. The full density matrix is represented by $\rho_{\textrm{tot}}(t)$. It is usually assumed that the initial state is an uncorrelated state i.e., at $t=0$, the system and the environment have separate states in the form,
\beq
 \rho_{\textrm{tot}}(0)=\rho_\S(0)\otimes\rho_\mathcal{E}~,
 \eeq
 where $\rho_\S$ and $\rho_\mathcal{E}$ are the system and the environment density operators, respectively. This may be the case if the system and the environment have not interacted at previous times or if the correlations between them are short-lived. 
We also assume that the initial state of the environment is thermal, meaning that it is described by the Gibbs state,  
\beq
\rho_\mathcal{E}=\frac{\exp (-\beta H_\mathcal{E})}{\textrm{Tr}[\exp(-\beta H_\mathcal{E})]}~,
\eeq
where $\beta$ is the inverse temperature, and $\rho_\mathcal{E}$ satisfies the stationary condition of the environment,
\beq
[\rho_\mathcal{E},H_\mathcal{E}]=0~.
\eeq
Because we are working in the weak-coupling limit, we can assume that the system and the environment are uncorrelated during the time evolution. On the other hand, we assume that the environment is so large that it is hardly influenced through interaction with the system. In this condition, the interaction between the system and the environment is such that the influence of the system on the environment is small, and one can assume that the time-scales of correlations of the environment, $\tau_{\mathcal{E}}$, is much smaller than the typical system time-scale 
$\tau_{\mathcal{S}}$. Due to this requirement, the environment is assumed to be in equilibrium in such a way that it is essentially unaffected by its coupling to the system. Therefore, the environment is unchanged in time, and the dynamics of the system is not affected by its coupling to the environment at earlier times. In this approximation, the state of the total system at time $t$ is approximately factorized by a tensor product,
\beq
\hat{\rho}_{\textrm{tot}}(t) \approx \hat{\rho}_\mathcal{S}(t) \otimes\hat{\rho}_\mathcal{E}+\mathcal{O}(g)~.
\eeq 
As it was mentioned, this is the \textit{Born approximation} \cite{Bruus}. 
If the initial state of the overall system is the state of the product, then it is assumed that the evolved state at a particular time $t$ is in the same product form.
The dynamics of the total system is govern by the Von-Neumann equation,
\begin{equation}
\label{vonNeumann}
\frac{d}{dt}\rho_{\textrm{tot}}(t)=-i[H_{\textrm{int}}(t),\rho_{\textrm{tot}}(t)]~.
\end{equation}
By integrating equation \eqref{vonNeumann}, inserting it once again in 
equation \eqref{vonNeumann}, and taking the partial trace on the environment degrees of freedom, we obtain an 
integro-differential equation for the reduced density matrix of the system 
\beq
\label{masterequ}
\frac{d}{dt}\rho_\S(t)=-\int_{0}^t dt' 
\,\Tr_\mathcal{E}[H_{\textrm{int}}(t),[H_{\textrm{int}}(t'),\rho_\S(t')\otimes\rho_\mathcal{E}]]+\mathcal{O}(g^3)~.
\eeq
To better understand the Markov approximation, we decompose the interaction Hamiltonian in the interaction picture and represent it in 
the following general form:
\beq
\label{decompositionIntH}
H_{\textrm{int}}(t)=\sum_\beta \hat{S}_\beta(t)\otimes \hat{E}_\beta(t)~,
\eeq
where $\hat{S}_\beta(t)$ and $\hat{E}_\beta(t)$ are the Hermitian operators associated with the system and the environment, respectively. 
The operators $\hat{S}_\beta(t)$ and $\hat{E}_\beta(t)$ can be expanded in terms of the creation and annihilation operators of the system and environment degrees of freedom.
We also assume that $\hat{S}_\beta(t)$ and $\hat{E}_\beta(t)$ commute because they are associated with different particles. In this sense, after changing the variable $s=t-t'$ and inserting equation \eqref{decompositionIntH} in equation \eqref{masterequ}, and
after some straightforward algebra, we get
\begin{eqnarray}
\label{eqmaster2}
\frac{d}{dt}\rho_\S(t)=&-\sum_{\beta,\beta'}\int_0^t ds
\left[\mathcal{E}_{\beta\beta'}(s)[ \hat{S}_\beta(t), \hat{S}_{\beta'}
(t-s)\rho_\S(t-s)]+\textrm{h.c.}\right] 
+\mathcal{O}(g^3)~,
\end{eqnarray}
where $\mathcal{E}_{\beta\beta'}(s)=\langle 
\hat{E}_\beta(s)\hat{E}_{\beta'}(0)\rangle_\mathcal{E}=Tr[\hat{E}_\beta(s)\hat{E}_{\beta'}(0)\rho_\mathcal{E}]$ is defined as the environment correlation function. Another fundamental approximation is to assume that the environment has a very 
short correlation time, $\tau_\mathcal{E}$, with correlation function that decays as 
$\abs{\mathcal{E}_{\beta\beta'}(s)}\sim e^{-s/\tau_\mathcal{E}}$. 
As previously stated, the environment correlation function $\mathcal{E}_{\beta\beta'}(s)$ decays sufficiently fast over a time $\tau_\mathcal{S}$. 
In fact, in the weak coupling limit, one can set $\tau_\mathcal{E}\ll \tau_\S$, in the sense that the system will 
relax slowly compared to the evolution of the environment correlation functions. 
If we now calculate the integral in equation \eqref{eqmaster2} for a 
sufficiently large time $t^*\gg\tau_\mathcal{E}$, such that $t^*$ is still much smaller 
than the time $\tau_\S$ at which the state of the system in the interaction picture 
changes considerably, then we can safely replace $\rho_\S(t-s)$ with 
$\rho_\S(t)$ in the same equation because the dynamics of $\rho_\S(t)$ is much
slower than the decay of the correlation function $\mathcal{E}_{\beta\beta'}(s)$. This replacement makes the master equation local in time. 
For the same reason, we can extend the upper limit of the integral to the infinity, since the added part will give a 
negligible contribution. This is known as the \textit{Markov approximation} implying that the memory effect of the
environment is negligible. The equation \eqref{eqmaster2} is known as the Bloch-Redfield equation \cite{Breuer,Lidar,Vega}. A simplified version of the Bloch-Redfield equation is obtained by taking the secular approximation in which the oscillating terms, of the form $e^{i(\omega'-\omega)t}$, are neglected. 
If there exist values of $\omega'$ and 
$\omega$ in equation \eqref{eqmaster2} being coarse-grained in time as from \cite{Cattaneo}
\begin{equation}
\label{eqn:secularAppCond}
\exists\,t^*\textnormal{ such that }\abs{\omega'-\omega}^{-1}\ll t^*\ll 
\tau_\S=O(g^{-2})~,
\end{equation}
then the terms in equation \eqref{eqmaster2} oscillating with frequency 
$\omega'-\omega$ will not give any significant contribution to the system evolution. That is because by integrating equation~\eqref{eqmaster2} for a time $t^*$ such that 
$\abs{\omega'-\omega}^{-1}\ll t^*\ll \tau_\S$ the fast-oscillating quantities 
vanish. Neglecting the fast-oscillating terms in the interaction picture is usually 
referred to as the {\it secular approximation} \cite{Cattaneo,Breuer,Lidar,Vega}. 
This approximation ensures that the master equation is in 
the Gorini-Kossakowski-Sudarshan-Lindblad (GKLS) form \cite{Breuer,Lidar,Vega}, and it 
therefore generates a dynamical semigroup, i.e., a perfectly Markovian evolution. 

\subsubsection{Dynamical map}

It is also worth sketching briefly the concept of dynamical maps and their relation
to the theory of open quantum systems. Using the quantum dynamical maps and their semigroup property one can find the sufficient condition for the Markovian dynamics. 

In general, the time evolution of the density matrix can be written as \cite{ Vega,Dariusz}
\beq
\rho_\S(t)=\exp(\mathcal{L}(t))\rho_\S(0)\equiv \Phi(t)\rho_\S(0)~,
\eeq
where $\mathcal{L}$ is the super-operator and $\Phi$ is known as the {\it dynamical map} that maps the density matrix from $t=0$
to its form at time $t$. The dynamical map is trace-preserving and also is completely positive, mapping a positive density matrix 
onto another positive density matrix. Formally, we can express the consequences of the 
Markovian approximation on the dynamical map as \cite{ Vega,Dariusz}
\beq
\Phi(t_1)\Phi(t_2)=\Phi(t_1+t_2) \:\:\:\:\:\: t_1,\: t_2 \geqslant 0~.
\eeq
In this situation, $\Phi$ forms a continuous linear dynamical semigroup.
Applying the Markovian approximation, the dynamics of an open quantum system is given by the 
following local master equation
\beq \label{master01}
\frac{d}{dt}\rho_\S(t)=\mathcal{L}(t)\rho_\S(t)~.
\eeq
Starting from this equation and using the dynamical semigroup approach one can 
derive the Lindblad master equation \cite{ Vega,Dariusz}. The non-Markovian generalization of \eqref{master01} is the following non-local equation: 
\beq \label{master02}
\frac{d}{dt}\rho_\S(t)=\int_0^t ds \mathcal{K}_\S(t-s)\rho_\S(s)~,
\eeq
where $\mathcal{K}_\S$ is a {\it memory kernel},  
which simply means that the rate of change of the state $\rho_\S(t)$ at time $t$ depends on its history (starting at t = 0). 


\subsection{ Quantum Boltzmann equation }

The master equation approach is suitable for applications in condensed matter physics and quantum technologies. However, to study the dynamics of the systems on an expanding background we must use other techniques. 
Here, we take the approach of the so-called QBE instead \cite{Kosowsky:1994cy,Bavarsad:2009hm,Bartolo:2018igk,Bartolo:2019eac,Bartolo:2020htk,Fidler:2017pkg} (see \cite{Burgess:2014eoa,Burgess:2015ajz,Boyanovsky:2015tba,Boyanovsky:2015jen,Nelson:2016kjm,Hollowood:2017bil,Martin:2018zbe,Shandera:2017qkg} for the application of OQS to the inflation models).
We aim to apply our non-Markovian formalism to a system of soft gravitons coupled to a background of decoupled ultra-relativistic fermions. 
However, our results are quite general and, in principle, can be applied to similar systems, such as CMB radiation, cosmic neutrinos, and dark matter. 
It should be noted that this approach is essentially similar to the method of the master equation described in the previous section, except that taking the trace over the environment states is replaced by taking expectation values over the environment's creation and annihilation operators. Therefore, physical processes can be calculated microscopically with the help of field theory techniques. In the following, we formulate QBE under the above conditions.

\subsubsection{ QBE with Born-Markov approximation but without secular approximation  }

To find the quantum Boltzmann equation, we start with the time evolution of the number operator associated with the system's degrees of freedom given in the following form \cite{Kosowsky:1994cy,Bavarsad:2009hm,Bartolo:2018igk,Bartolo:2019eac,Bartolo:2020htk,Fidler:2017pkg}:
\beq \label{boltzmanneq0}
\frac{d}{dt}\hat{\mathcal{N}}^\mathcal{S}_{ij}(\k,t)=i\left[H^0_{\textrm{int}}(t),\hat{\mathcal{N}}^S_{ij}(\k,t)\right]- \int_{0}^{t}ds\, \left\{\left[H^0_{\textrm{int}}(t),\left[H^0_{\textrm{int}}(t-s),\hat{\mathcal{N}}^S_{ij}(\k,t-s)\right]\right]\right\}~.
\eeq 
The knowledge about scattering processes is encoded in the S-matrix element. 
It is essential to note that, for a given process, 
the effective interaction Hamiltonian $H^0_{\textrm{int}}(t)$ is defined using the n-th order S-matrix \cite{Kosowsky:1994cy}
\beq 
\label{smatrix}
S^{(n)}=-i\int dtH^0_{\textrm{int}}(t)~,
\eeq 
where the superscript $0$ indicates that the interaction Hamiltonian is a functional of the free field. A discussed in detail in the \cite{Kosowsky:1994cy,Bavarsad:2009hm,Bartolo:2018igk,Bartolo:2019eac,Bartolo:2020htk,Fidler:2017pkg}, $H^0_{\textrm{int}}(t)$ describes physical processes such as scattering and decay phenomena. 
The superscript $n$ in \eqref{smatrix} shows
the number of vertices in the corresponding Feynman
diagrams of such process. Each vertex corresponds to
the fundamental interaction Hamiltonian $H_I(g)$ in which
$g$ denotes a general dimensionless coupling constant. Accordingly, it should be noted that $H^0_{\textrm{int}}(t)$ is different from the fundamental interaction Hamiltonian.
 Moreover, 
$\hat{\mathcal{N}}^\mathcal{S}_{ij}(\k,t)$ is the number operator of the system defined as
\beq 
\hat{\mathcal{N}}^\mathcal{S}_{ij}(\k,t)=a_i^{\dag}(\k,t)a_j(\k,t)~,
\eeq 
where $a_i$ and $a^\dag_i$ are the creation and annihilation operators respectively, associated with the system's degrees of freedom. 
The number operator is related to the system's density matrix after taking the expectation value in the following form
\beq \label{trace}
\l<\hat{\mathcal{N}}^\mathcal{S}_{ij}(\k,t)\r>= \textrm{tr}[\hat{\rho}^{(\mathcal{S})}\hat{\mathcal{N}}^\mathcal{S}_{ij}(\k,t)]=(2\pi)^3\delta^3(0)2k^0\rho^\mathcal{S}_{ij}(\k,t)~,
\eeq 
where trace over the continuum of states is defined in Appendix A. 
It is difficult to solve the integro-differential equation \eqref{boltzmanneq0} because it is non-local in time, as $\mathcal{N}^{\mathcal{S}}_{ij}$ still depends upon the entire history of the process, and the integration runs over time. This can be reduced to an equation local in time if there is a clear separation of time-scales. 
To resolve this problem, we make the Markov approximation, in which the time-scale of the environment is taken to be much shorter than the time-scale of the system so that the memory effects of the environment are negligible in the long run. In order to perform this approximation, we replace $\hat{\mathcal{N}}^\mathcal{S}_{ij}(\k,t-s)$ by $\hat{\mathcal{N}}^\mathcal{S}_{ij}(\k,t)$ due to its slow evolution. In this way, we separate the time-scales into microscopic time-scale $t_{\textrm{mic}}$, quantifying the interaction time-scale of individual particles, and mesoscopic time-scale $t_{\textrm{mes}}$, quantifying the time-scale on which the whole macroscopic system evolves. Therefore, in the Born-Markov approximation, 
the time evolution of this system is given by the following master equation \cite{Kosowsky:1994cy,Bavarsad:2009hm,Bartolo:2018igk,Bartolo:2019eac,Bartolo:2020htk,Fidler:2017pkg}:
\bea \label{boltzmanneq1}
\frac{d}{dt_{\textrm{mes}}}\hat{\mathcal{N}}^\mathcal{S}_{ij}(\k,t_{\textrm{mes}})&=&i\left[H^0_{\textrm{int}}(t_{\textrm{mes}}),\hat{\mathcal{N}}^\mathcal{S}_{ij}(\k,t_{\textrm{mes}})\right]
\nonumber \\&-&
 \int_{0}^{t_{\textrm{mes}}}dt_{\textrm{mic}}\left[H^0_{\textrm{int}}(t_{\textrm{mes}}),\left[H^0_{\textrm{int}}(t_{\textrm{mes}}-t_{\textrm{mic}}),\hat{\mathcal{N}}^S_{ij}(\k,t_{\textrm{mes}})\right]\right]~.
\eea  
As explained in the previous section, the time integration can be extended to infinity due to the Born-Markov approximation. Now, after extending the upper limit of the integral involving $\tmes$ to the infinity
and taking the
expectation value of both sides of \eqref{boltzmanneq1}, we find the final form of the quantum Boltzmann equation as 
\bea \label{boltzmanneq2}
(2\pi)^3\delta^3(0)2k^0\frac{d}{dt_{\textrm{mes}}}\rho^\mathcal{S}_{ij}(\k,\x,t_{\textrm{mes}})&=&
i\l<\left[H^0_{\textrm{int}}(t_{\textrm{mes}}),\hat{\mathcal{N}}^S_{ij}(\k,t_{\textrm{mes}})\right]\r>_\textrm{c}
\nonumber \\&-&
 \int_{0}^{\infty}dt_{\textrm{mic}}\l<\left[H^0_{\textrm{int}}(\tmes),\left[H^0_{\textrm{int}}(-t_{\textrm{mic}}),\hat{\mathcal{N}}^\mathcal{S}_{ij}(\k,t_{\textrm{mes}})\right]\right]\r>_\textrm{c}~,
\eea
where the subscript $c$ labels what we consider the connected part of the correlation functions. 
Upon taking the expectation values, the macroscopic properties of the system (for example, GWs) emerge. 
 It is also assumed that the process obeys the time-reversal symmetry. The operation of time-reversal interchanges initial and final states with identical positions but opposite momenta. For the scattering processes such as Compton scattering (the dominant interaction for CMB photons), the S-matrix is invariant under the interchange of initial and final states through which, the interaction Hamiltonian defined by equation \eqref{smatrix} is invariant under time-reversal. Therefore, we have 
$H^0_{\textrm{int}}(-t_{\textrm{mic}})=H^0_{\textrm{int}}(t_{\textrm{mic}})$ under which, the Boltzmann equation transforms in the following form \cite{ Kosowsky:1994cy,Fidler:2017pkg}:
 \bea \label{boltzmanneq30}
(2\pi)^3\delta^3(0)2k^0\frac{d}{dt_{\textrm{mes}}}\rho^\mathcal{S}_{ij}(\k,\x,t_{\textrm{mes}})&=&
i\l<\left[H^0_{\textrm{int}}(t_{\textrm{mes}}),\hat{\mathcal{N}}^\mathcal{S}_{ij}(\k,t_{\textrm{mes}})\right]\r>_\textrm{c}
\nonumber \\&&\!\!\!\!\!
-
\frac{1}{2} \int_{-\infty}^{\infty}dt_{\textrm{mic}}\left<\left[H^0_{\textrm{int}}(\tmes),\left[H^0_{\textrm{int}}(t_{\textrm{mic}}),\hat{\mathcal{N}}^\mathcal{S}_{ij}(\k,t_{\textrm{mes}})\right]\right]\right>_\textrm{c}~,
\eea
the equation that deals with the 
reversible scattering processes.
 The first term on the right side of \eqref{boltzmanneq2} is known as the forward scattering term, and the second term is the usual collision term \cite{
 Kosowsky:1994cy,Fidler:2017pkg}. It is worth emphasizing that in this expression $H^0_{\textrm{int}}(t_{\textrm{mic}})$ is also dependent on 
 $\tmes$, although we have not shown it explicitly. 
  
Here, it is worth emphasizing that in the above QBE there is no need to consider the secular approximation. In fact, as shown in \cite{Kosowsky:1994cy,Bavarsad:2009hm,Bartolo:2018igk,Bartolo:2019eac,Bartolo:2020htk,Fidler:2017pkg}, for the physical processes described by the interaction Hamiltonian, energy conservation is obtained naturally after taking the integration over $\tmic$ in the interaction picture.

 \subsubsection{ Extension to Markovian irreversible processes}

There are a variety of examples of irreversible processes in the early universe.
In this work, we finally intend to explain the damping of GWs by an environment containing decoupled ultra-relativistic fermions using the QBE. We assume that fermions are decoupled before re-entering the horizon during radiation dominance. Due to its dissipative nature, this process is considered an irreversible process. If we want to use the QBE to explain this damping effect, then we first need to identify the interaction Hamiltonian and essentially the microscopic process that leads to this irreversible phenomenon. 
The question of how macroscopic irreversibility emerges from microscopic processes has always been a fundamental question. 
The root of this problem is that we still do not know exactly how to reconcile the second law of thermodynamics with its intrinsic arrow of time, with the microscopic time-reversible dynamical equations.
On one hand, the process of GW damping is an irreversible phenomenon due to its dissipative nature, and on the other hand, it has a micro-reversibility property on microscopic scales  (see \cite{Agarwal} for the discussion on the relation of the detailed-balance with the property of micro-reversibility of the underlying microscopic dynamics). In order to use the QBE to explain this phenomenon, we must first generalize it to irreversible phenomena. 

In our formalism, the detailed balance condition or micro-reversibility is fulfilled for microscopic processes. 
Hence, in all the processes that we assume, the amplitudes of the initial to final and final to initial reactions are equal. In other words, the phenomenon emerging on the macroscopic scale is not invariant under time reversal, whereas we assume that the microscopic interactions are invariant under time reversal. 
The emergent irreversible process must occur on macroscopic scales. For this purpose, we assume that the system is in contact with an environment in a steady-state. Under these conditions, the generalization of the equation to irreversibility conditions will be perform as follows. 

For a microscopic irreversible process, like absorption or emission 
the effective interaction Hamiltonian doesn't satisfy the relation $H^0_{\textrm{int}}(-t_{\textrm{mic}})=H^0_{\textrm{int}}(t_{\textrm{mic}})$.
It is worth emphasizing once again that $H^0_{\textrm{int}}(-t_{\textrm{mic}})$ actually describes the physical process and is different from the fundamental interaction Hamiltonian.
Therefore, one cannot use the equation \eqref{boltzmanneq30} to deal with such processes. 
The effective interaction Hamiltonian for a specific irreversible process is written in terms of the creation and the annihilation operators. For such a process, the action of the time-reversal transformation on $H^0_{\textrm{int}}$ is equivalent to the action of Hermitian conjugation. 
We therefore generalize the equation \eqref{boltzmanneq2} by replacing $H^0_{\textrm{int}}(t_{\textrm{mic}})$ by the following Hermitian
combination:
\beq \label{H0dag}
 H^0_{\textrm{int}}(-t_{\textrm{mic}})\rightarrow  H^{0\dag}_{\textrm{int}}(t_{\textrm{mic}})~.
 \eeq
By doing this, the collision term is modified as follows,
  \bea \label{boltzmanneq4}
  -
\int_{0}^{\infty}dt_{\textrm{mic}}\left<\left[H^0_{\textrm{int}}(\tmes),\left[ H^{0\dag}_{\textrm{int}}(t_{\textrm{mic}}) ,\hat{\mathcal{N}}^{\mathcal{S}}_{ij}(\k,t_{\textrm{mes}})\right]\right]\right>_\textrm{c}~,
\eea
in which the forward scattering term automatically vanishes for such emission or absorption processes. Using this equation we will be able to compute the time evolution of density matrix for irreversible phenomena such as decaying or absorption process. In the condition that $H^{0}_{\textrm{int}}(t_{\textrm{mic}})=H^{0}_{\textrm{int}}(-t_{\textrm{mic}})$, we will reproduce the conventional collision term.

\subsubsection{Extension to non-Markovian (ir)reversible processes}

In the remainder of this section, we discuss deviations from the Markovian approximation. As discussed above, in most situations the non-Markovianity appears to be relevant for time-scales smaller than, or of the order of, the environment correlation time $\tau_{\mathcal{E}}$.  

For the case of the interaction of SBGW with a background of ultra-relativistic fermion bath and when one ignores back-reaction effects, the time evolution of the intensity of GW can be casted in equation \eqref{boltzmanneq4}. However, for a more realistic situation when the back-reaction effects, are important one cannot use this equation. In general, as already emphasized, the underlying assumption of weak system-environment coupling to an essentially unchanging, memoryless environment is not always fulfilled in many situations of physical interest, and significantly non-Markovian dynamics may arise. If memory effects in the environment are substantial, then the evolution of the reduced density matrix will depend on the past history of the system and the environment. In this condition, 
 information can also flow back from the environment to the system, resulting in a back-reaction effect of the environment. 
The microscopic description of non-Markovian dynamics is much more complicated than the Markovian one. The precise details of the non-Markovian dynamics has still not fully worked out, partly because of the complexity behind such phenomena. 
Here, we will generalize the QBE as a new tool to deal with the non-Markovian processes. The generalized QBE is derived as follows: First, it is important to note that we still consider the Born approximation. In addition, 
 as stated, we are interested in the absorption processes in which the time-reversal symmetry is satisfied microscopically, but the process is irreversible macroscopically. 
 For this case, time-local master equations are no longer applicable, and one has to instead solve integro-differential equations. We also need to replace the interaction Hamiltonian with the relation \eqref{H0dag} so that the QBE can properly describe an absorption process. The equation we will ultimately work with to describe an irriversal and non-Markovian process is as below 
\bea \label{boltzmanneq3}
(2\pi)^3\delta^3(0)2k^0\frac{d}{dt_{\textrm{mes}}}\rho^\mathcal{S}_{ij}(\k,\x,t_{\textrm{mes}})= D_{ij}[\rho^S(\k,\x,t_{\textrm{mes}})]~,
\eea
where $D_{ij}[\rho^S(\k,\x,t_{\textrm{mes}})]$ is the ``dissipator" that is given by
\bea
 D_{ij}[\rho^S(\k,\x,t_{\textrm{mes}})]= -\int_{0}^{t_{\textrm{mes}}}dt_{\textrm{mic}}\left<\left[H^0_{\textrm{int}}(\tmes),\left[ H^{0\dag}_{\textrm{int}}(t_{\textrm{mic}}) ,\hat{\mathcal{N}}^S_{ij}(\k,t_{\textrm{mes}}-t_{\textrm{mic}})\right]\right]\right>_\textrm{c}~.
\eea
While $\hat{\mathcal{N}}^S_{ij}$ is a non-local operator in time, we have kept the interaction Hamiltonian local in time for simplicity.
In the following, we calculate the  dissipative term $D_{ij}[\rho^S(\k,t_{\textrm{mes}})]$ for the graviton absorption process by a decoupled ultra-relativistic fermions.


\section{Application: absorption of soft gravitons by a ultra-relativistic fermion bath}

As anticipated, in this work, we apply the non-Markovian QBE to study the damping phenomena due to coupling with ultra-relativistic fermions. Before, we characterize the system, the environment and the system-environment interaction Hamiltonian. 

\subsection{Open quantum system components}

As mentioned above, the SBGW absorption process can be studied through the approach of an open system interacting with the environment. The system is SGWB propagating in an environment containing decoupled ultra-relativistic fermions. In the following, we will consider an example that such fermions can be considered the same as decoupled neutrinos. We generally assumed that both the system and the environment are affected by each other. This interaction is described by $H^0_{\textrm{int}}(t)$. We then outline the system, the environment, and the interaction Hamiltonian.  We also emphasize once again that our system and the environment are on an expanding space-time background. In this condition, we must consider a new time-scale that is the inverse of the Hubble parameter $H^{-1}$. We will discuss this new time-scale further below.

\subsubsection{System}

We consider a quantum system of soft graviton degrees of freedom that is affected by its coupling to the environment. The SGWB field $h_{\mu\nu}$ is given by assuming 
 the weak-field limit and expanding the metric around Minkowski space-time as follows:
 \beq  
g_{\mu\nu}=\eta_{\mu\nu}+\kappa h_{\mu\nu}~,
\eeq 
where $\kappa=\sqrt{16\pi G}$. In the following, we will straightforwardly generalize this metric to the FRW background. The dynamics of free gravitons in the transverse-traceless gauge is given by the following Lagrangian density:
\beq \label{hgraviton}
\mathcal{L}_{\textrm{g}}=\frac{1}{4} \l [ \dot{h}_{\mu\nu}\dot{h}^{\mu\nu} + \partial_\lambda h_{\mu\nu}\partial^\lambda h^{\mu\nu}   \r ]~.
\eeq
The associated quantum field is decomposed as
\beq\label{quant}
h_{\mu\nu}(x)=h^+_{\mu\nu}(x)+h^-_{\mu\nu}(x)~,
\eeq 
where $h^-_{\mu\nu}(x)$ and $h^+_{\mu\nu}(x)$ are linear in graviton creation and annihilation operators respectively. Fourier transforms of the fields are expressed by the following conventions:
\bea \label{h1}
h^+_{\mu\nu}(x)&=&
\int d\p\sum_{s=+,\times}a_s(\mathbf{p},t)\,h^{s}_{\mu\nu}(p)\,e^{-i(p^0 t-\p\cdot \x)}~, \\ \label{h2}
h^-_{\mu\nu}(x)&=&
\int d\p\sum_{s=+,\times} a^{\dag}_{s'}(\mathbf{p},t)\,h^{s\,\ast}_{\mu\nu}(p)\,e^{i(p^0 t-\p\cdot \x)}~,
\eea
where $a_s$ and $a^\dag_s$ are the graviton annihilation and creation operators, and the abbreviation $d\p$ is defined as
\beq 
d\p=\frac{d^3p}{(2\pi)^3}\frac{1}{2p^0}~,
\eeq
and $h^{(r)}_{\mu\nu}$ are the polarization tensors with the following well-known properties:
\beq
h^{s}_{\mu\nu}(p)p^{\mu}=0~,\:\:\:\:\:\:\:\:h_{\mu}^{\mu}(p)=0~,\:\:\:\:\:\:\:\:h^{s}_{\mu\nu}(p)
\left(h^{s'\,\mu\nu}(p)\right)^{\ast}= \delta^{ss'}~ \label{canonical}.
\eeq
Note that in \eqref{h1} and \eqref{h2}, we have not separated the microscopic and mesoscopic times to avoid confusion. We assume that the time appeared in the exponential function is a microscopic time, while the annihilation and creation operators can generally be a function of both times. 
It is also convenient to represent the polarization tensor $h^{(s)\,\mu\nu}$ in terms of a direct product of unit spin polarization vectors,
\beq
h^{s}_{\mu\nu}(p)=e^{s}_{\mu}(p)e^{s}_{\nu}(p)~,\:\:\:\:\:\:\:e^{s}_{\mu}(p)p^{\mu}=0 ~,\:\:\:\:\:\:\:
  \left[ e^{s}_{\mu}(p) \left(e^{s'\,\mu}(p)\right)^{\ast} \right]^2=  \delta^{ss'}~.
\eeq
In general, $a_s$ and $a^\dag_s$ are assumed to be time-dependent. In Appendix A, we will present a general discussion about their time-dependent commutation relation and the calculation of their expectation values. 
As shown in this appendix, for equal times, the canonical commutation relations are given by
\beq \label{commutationa}
\left[a_s(p,t),a^{\dag}_{s'}(p',t)\right]=(2\pi)^32p^0\delta^3(\p-\p')\delta_{ss'}~.
\eeq 
The graviton density operator is presented in the following form \cite{Bartolo:2018igk}:
\beq \label{rhohatg}
\hat{\rho}^{(\textrm{g})}(\x,t)=\int \frac{d^3p}{(2\pi)^3}\rho^{(\textrm{g})}_{ij}(\x,t)a^{\dag}_i(\p,t)a_j(\p,t)~,
\eeq 
where the polarization matrices $\rho^{(\textrm{g})}$  for a system of gravitons have the following form:
\beq
\rho^{(\textrm{g})}(\x,t)=\frac{1}{2}\begin{pmatrix} I^{(\textrm{g})}(\x,t)+Q^{(\textrm{g})}(\x,t) & U^{(\textrm{g})}(\x,t)-iV^{(\textrm{g})}(\x,t)\\ U^{(\textrm{g})}(\x,t)+iV^{(\textrm{g})}(\x,t) &  I^{(\textrm{g})}(\x,t)-Q^{(\textrm{g})}(\x,t)\end{pmatrix}~,
\eeq 
where $I^{(\textrm{g})}$ denotes the radiation intensity, $Q^{(\textrm{g})}$ and $U^{(\textrm{g})}$ parameterize the linear polarization, and $V^{(\textrm{g})}$ is the circular polarization.  Among these parameters, $I^{(\textrm{g})}$ is always positive, while the other three parameters can have either sign. The Stokes parameters for monochromatic plane GWs are defined by \cite{Bartolo:2018igk}
\bea
I^{(\textrm{g})}&=&\l(h^+\r)^2+\l(h^\times\r)^2~,\\
Q^{(\textrm{g})}&=&\l(h^+\r)^2-\l(h^\times\r)^2~,\\
U^{(\textrm{g})}&=&2\cos\alpha\, h^+h^\times~,\\
V^{(\textrm{g})}&=&2\sin\alpha \, h^+h^\times~,
\eea
where $h^+$ and $h^\times$ are the complex amplitude for two GW modes, and $\alpha$ represents the difference of the phases of the $h^+$ and $h^\times$ modes. $Q^{(\textrm{g})}$ measures the difference between polarization modes, $U^{(\textrm{g})}$ and $V^{(\textrm{g})}$ measure the phase-dependence of modes. The condition of $Q^{(\textrm{g})}=U^{(\textrm{g})}=V^{(\textrm{g})}=0$ is associated with an unpolarized GW. In this case, the GW is composed of incoherent modes with random polarization angles.
The phase of the $h^+$ and $h^\times$ modes will change over the coherence time, which is taken much greater than the period of a quasi-monocratic wave. 
 However, in the condition that the phase of modes remains essentially constant over a duration shorter than the coherence time, the GW becomes polarized.

 \subsubsection{Environment}   
    
The environment contains decoupled ultra-relativistic fermionic degrees of freedom and behaves like a thermal fermion bath. It is also assumed that the environment (i) is expanding, (ii) is in thermal equilibrium, 
(iii) notices that the system is interacting with it. However, it cannot relax back to its equilibrium in a short time due to the expansion of space-time. Figuratively, it means that  the environment has memory. We will incorporate this kind of memory effect due to the expanding space-time in our calculation. 

Fermions are generally described by the spinor field $\psi_{f}$ that is decomposed as $\psi_{f}(x)=\psi^+_{f}(x)+\psi^-_{f}(x)$ such as
\bea \label{psi1}
\bar{\psi}^-_{f}(x)&=&\int d\q\sum_r  b^{\dag}_r(q,t)\bar{u}_r(q)e^{i(q^0 t-\q\cdot \x)}~,
\\ \label{psi2}
\psi^+_{f}(x)&=&\int d\q\sum_r 
b_r(q,t)u_r(q)e^{-i(q^0 t-\q\cdot \x)}~,
\eea
where $u_r$ is the Dirac spinor, with spin index $r =1,2$, $b_r$, and $b^\dag_r$ are fermion creation and annihilation operators, respectively, and
\beq 
d\q=\frac{d^3q}{(2\pi)^3}~.
\eeq 
The creation and the annihilation operators of fermions obey 
the following equal-time canonical anti-commutation relation 
\beq
\left\{b_r(q,t),b^\dag_{r'}(q',t)\right\}=(2\pi)^3\delta^3(\q-\q')\delta_{rr'}~.
\eeq
 Using these operators, we can define the fermionic density operators in the form,   
\beq \label{rhohatf}
\hat{\rho}^{(f)}(\x,t)=\int \frac{d^3q}{(2\pi)^3}\rho^{(f)}_{ij}(\x,\q, t)b^{\dag}_i(\q,t)b_j(\q,t)~,
\eeq 
where for a system composed of unpolarized fermions, $\rho^{(f)}_{ij}$ is given by
\beq
\rho^{(\textrm{f})}(\x,t)=\frac{1}{2}\begin{pmatrix} n^{(f)}(\x,t) & 0\\ 0&  n^{(f)}(\x,t)\end{pmatrix}~,
\eeq 
where, $n^{(f)} $ denotes the intensity of the fermions.

\subsubsection{Interaction Hamiltonian}

The gravitons interact with an environment composed of the
ultra-relativistic fermions that causes a damping effect. As we mentioned before, the QBE is a new powerful tool to obtain a classical macroscopic description of this effect emerging from a microscopic absorption. Voronov has calculated the scattering amplitude of the gravitons with fermions many years ago \cite{Voronov:1973kga}. Here, we use his results to calculate the absorption (emission) of gravitons by (from) fermions. 
Using the S-matrix element, we can write the effective interaction Hamiltonian describing the absorption of graviton by fermions of thermal bath as in the following form:
\bea \label{hgnu1}
H^0_{\textrm{int}}(t)= -\frac{i}{2}\kappa\int d^3 x\,
h^+_{\mu\lambda}\bar{\psi}^-\gamma^{\lambda}\partial^{\mu}\psi^+
~.
\eea 
Inserting the Fourier transforms \eqref{h1}, \eqref{psi1}, and \eqref{psi2}, into \eqref{hgnu1} we get
\bea \label{hgnu2}
H^0_{\textrm{int}}(t)&= &-\frac{i}{2}\kappa\int d^3 x  d\p d\q d\q'\sum_{s,r,r'}
h^s_{\mu\lambda}(p) \bar{u}_{r'}(q')(-iq^{\mu})\gamma^{\lambda}u_r(q)
\nonumber \\ && ~~~~~~~~~~~\times e^{i(-p^0+q'^0-q^0)t}e^{-i(-\p+\q'-\q)\cdot \x}
a_s(p,t)b^{\dag}_{r'}(q',t)b_{r}(q,t)
~.
\eea 
Integrating over $\x$ gives
\bea \label{hgnu2}
H^0_{\textrm{int}}(t)&= &-\frac{i}{2}\kappa\int  d\p d\q d\q'\sum_{s,r,r'}
h^s_{\mu\lambda}(p) \bar{u}_{r'}(q')(-iq^{\mu})\gamma^{\lambda}u_r(q)(2\pi)^3\delta^3(\q'-\q-\p)
\nonumber \\ && ~~~~~~~~~~~\times 
e^{i(-p^0+q'^0-q^0)t}a_s(p,t)b^{\dag}_{r'}(q',t)b_{r}(q,t)
~.
\eea 
In this step, we deduce the behavior of \eqref{hgnu1} under time reversal. It is known that the covariant bilinear transforms under time reversal as \cite{Itzykson,Berestetskii}
\beq 
\bar{\psi}_a\gamma^{\lambda}\partial^{\lambda}\psi_b   \xrightarrow{\text{T}} \bar{\psi}_b\gamma^{\lambda}\partial^{\lambda}\psi_a~,
\eeq 
where $a,b=+,-$. Also, the GW field transforms as
\beq 
h^+_{\mu\nu}\xrightarrow{\text{T}} 
h^-_{\mu\nu}~.    
\eeq 
Therefore, we get 
\bea \label{hgnu2}
H^0_{\textrm{int}}(-t)&= &-\frac{1}{2}\kappa \sum_{s,r,r'}\int d^3 x  d\p d\q d\q'
h^{\ast s}_{\mu\lambda}(p) \bar{u}_{r}(q)q'^{\mu}\gamma^{\lambda}u_{r'}(q')
e^{-i(-p^0+q'^0-q^0)t} 
\nonumber \\ && ~~~~~~~~~~~\times
e^{-i(\p+\q-\q')\cdot \x}a^{\dag}_s(p,t)b^{\dag}_{r}(q,t)b_{r'}(q',t)
\nonumber \\ &  =&
H^{0\dag}_{\textrm{int}}(t)
~,
\eea 
which confirms our claim about the relation of time-reversal and complex conjugate operations on the interaction Hamiltonian. $H^{0\dag}_{\textrm{int}}(t)$ describes the inverse process, i.e., the graviton emission from the fermion of thermal bath. 
As explained in the previous section, the QBE involves both $H^{0}_{\textrm{int}}(t)$ and $H^{0\dag}_{\textrm{int}}(t)$ terms.
After this extension, the collision term will result in two processes that will be interpreted as the graviton absorption and emission processes. 
One can apply the same formalism to the interaction of gravitons with ultra-relativistic anti-fermions.



  At the end of this part, we briefly discuss the action of parity operator on interaction Hamiltonian \eqref{hgnu1}. In general, the covariant bilinear is transformed under a parity transformation as \cite{Itzykson,Berestetskii}
  \beq 
\bar{\psi}_a\gamma^{\lambda}\partial^{\lambda}\psi_b   \xrightarrow{\text{P}} \bar{\psi}_a\gamma^{\lambda}\partial^{\lambda}\psi_b~.
\eeq 
It is also shown that $h^{R}_{\mu\nu} \rightarrow  h^{L}_{\mu\nu}$ under a parity transformation \cite{Bartolo:2018qqn}, where $L$ and $R$ represent, respectively, the left-handed and the right-handed GW circular polarization. Therefore, the interaction Hamiltonian $h^{(s)}_{\mu\nu}\bar{\psi}_a\gamma^{\lambda}\partial^{\lambda}\psi_b$
is not totally invariant under a parity transformation.


\subsection{Calculation of the collision term }

In this part, we provide the general calculation of the collision or damping term. 
Working in a comoving frame, we substitute the interaction Hamiltonian \eqref{hgnu2} into the dissipative term on the right-hand side of equation \eqref{boltzmanneq3} and find
\bea  \label{scattering2}
D_{ij}[\rho^{\g}(\k,\x,t_{\textrm{mes}})]&=&-
\frac{\kappa^2}{4}\int_{0}^{\tmes}dt_{\textrm{mic}}\int d^3x d^3x'd\p_1d\q_1d\q'_1d\p_2d\q_2d\q'_2
\,e^{i(p^0_2+q^0_2-q'^0_2)\tmic}
\nonumber \\ && \!\!\!\!\!\! \times
\sum_{s_1,r_1,r'_1}\sum_{s_2,r_2,r'_2}
h^{s_1}_{\mu_1\lambda_1}(p_1) \bar{u}_{r'_1}(q'_1)q_1^{\mu_1}\gamma^{\lambda_1}u_{r_1}(q_1)
h^{\ast\, s_2}_{\mu_2\lambda_2}(p_2) \bar{u}_{r_2}(q_2)q'^{\mu_2}_2\gamma^{\lambda_2}u_{r'_2}(q'_2)
\nonumber \\ &\times&  e^{-i(\q'_1-\q_1-\p_1)\cdot\x_1}
e^{-i(\p_2+\q_2-\q'_2)\cdot\x_2}
\left<\left[a_{s_1}(p_1,\tmes)b^{\dag }_{r'_1}(q'_1,\tmes)b_{r_1}(q_1,\tmes)
\right.\right.\nonumber \\ && \left.\left. 
,\left[a^{\dag}_{s_2}(p_2,\tmic))b^{\dag }_{r_2}(q_2,\tmic))b_{r'_2}(q'_2,\tmic))
\right.\right.\right.\nonumber \\ &&~~~~~ \left.\left. \left.
,a^{\dag}_i(k,\tmes-\tmic))a_j(k,\tmes-\tmic)\right]\right]\right>_{\textrm{c}}~.
\eea 
Integrating over $\x$ and $\x'$  gives
\bea \label{scattering2}
D_{ij}[\rho^{\g}(\k,\x,t_{\textrm{mes}})]&=&-
\frac{\kappa^2}{4}\int d\p_1d\q_1d\q'_1d\p_2d\q_2d\q'_2\,(2\pi)^6\delta^3(\p_2+\q_2-\q'_2)\delta^3(\q'_1-\q_1-\p_1)
\,e^{i(p^0_2+q^0_2-q'^0_2)\tmic}
\nonumber \\ && \times \sum_{s_1,r_1,r'_1}\sum_{s_2,r_2,r'_2}
h^{s_1}_{\mu_1\lambda_1}(p_1) \bar{u}_{r'_1}(q'_1)q_1^{\mu_1}\gamma^{\lambda_1}u_{r_1}(q_1)
h^{\ast\, s_2}_{\mu_2\lambda_2}(p_2) \bar{u}_{r_2}(q_2)q'^{\mu_2}_2\gamma^{\lambda_2}u_{r'_2}(q'_2)
\nonumber \\ && \times
\left<\left[a_{s_1}(p_1,\tmes)b^{\dag }_{r'_1}(q'_1,\tmes)b_{r_1}(q_1,\tmes),\left[a^{\dag}_{s_2}(p_2, \tmic)b^{\dag }_{r_2}(q_2, \tmic)b_{r'_2}(q'_2, \tmic)
\right.\right.\right.\nonumber \\ && \left.\left.\left.\:\:\:\:\:\:\:\:\:\:
,a^{\dag}_i(k,\tmes-\tmic)a_j(k,\tmes-\tmic)\right]\right]\right>_{\textrm{c}}~.
\eea
Using the non-equal time expectation values computed in the Appendix A, we find the following expression:
\bea &&
\left<\left[a_{s_1}(p_1,\tmes)b^{\dag }_{r'_1}(q'_1,\tmes)b_{r_1}(q_1,\tmes),\left[a^{\dag}_{s_2}(p_2,\tmic)b^{\dag }_{r_2}(q_2,\tmic)b_{r'_2}(q'_2,\tmic),
\right.\right.\right.\nonumber \\ && \left.\left.\left.
a^{\dag}_i(k,\tmes-\tmic)a_j(k,\tmes-\tmic)\right]\right]\right>_{\textrm{c}}= \left<b^{\dag }_{r'_1}(q'_1,\tmes)b_{r_1}(q_1,\tmes)b^{\dag }_{r_2}(q_2,\tmic)b_{r'_2}(q'_2,\tmic)\right>_{\textrm{c}}
\nonumber \\ &\times& \l \{ \left<a_{s_1}(p_1,\tmes)a^{\dag}_{s_2}(p_2,\tmic)
a^{\dag}_i(k,\tmes-\tmic)a_j(k,\tmes-\tmic)\right>_{\textrm{c}}
\right.\nonumber \\ && \left.
~~~~~~~~~-\left<a_{s_1}(p_1,\tmes)
a^{\dag}_i(k,\tmes-\tmic)a_j(k,\tmes-\tmic)a^{\dag}_{s_2}(p_2,\tmic)\right>_{\textrm{c}}
 \r \}
\nonumber \\ &+&  
\left<b^{\dag }_{r_2}(q_2,\tmic)b_{r'_2}(q'_2,\tmic)b^{\dag }_{r'_1}(q'_1,\tmes)b_{r_1}(q_1,\tmes)\right>_{\textrm{c}}
\nonumber \\ && ~~~~~~~~~ \times
\l < 
a^{\dag}_i(k,\tmes-\tmic)a_j(k,\tmes-\tmic)a^{\dag}_{s_2}(p_2,\tmic)a_{s_1}(p_1,\tmes)
\r >_{\textrm{c}}
\nonumber \\ &\simeq& 
\frac{1}{2}(2\pi)^{12}4k^0p_1^0\delta^3(\q'_2-\q'_1)\delta^3(\q_1-\q_2)\delta^3(\k-\p_2)\delta^3(\p_1-\k)\delta_{r'_2r'_1}\delta_{r_1r_2}\delta_{js_2}\rho^{\g}_{s_1i}(\p_1,\x,\tmes-\tmic)
\nonumber \\ &\times& 
\l [n^{(f)}(\q_1,\tmes,\tmic)-n^{(f)}(\q'_1,\tmes,\tmic)\r]
~,
\eea 
where all terms quadratic in the system's density matrix have been dropped.  
Plugging this expression in \eqref{scattering2} and after taking integration over momenta, we find
\bea \label{scattering3}
D_{ij}[\rho^{\g}(\k,\x,t_{\textrm{mes}})]&=&-
\frac{\kappa^2}{8}\int_{0}^{\tmes}dt_{\textrm{mic}}\int d\q_1d\q'_1(2\pi)^6\delta^3(\k+\q_1-\q'_1)\delta^3(\q'_1-\q_1-\k )
e^{i(p^0_2+q^0_2-q'^0_2)\tmic}
\nonumber \\ & \times & \sum_{s_1,r_1,r'_1}q_1^{m_1}q_2^{m_2}
h^{s_1}_{m_1n_1}(\k) h^{\ast\, j}_{m_2n_2}(\k)\bar{u}_{r'_1}(\q'_1)\gamma^{n_1}u_{r_1}(\q_1)
 \bar{u}_{r_1}(\q_1)\gamma^{n_2}u_{r'_1}(\q'_1)
\nonumber \\ & \times &  
\rho^{\g}_{s_1i}(\k,\x,\tmes-\tmic)\l [n^{(f)}(|\q_1|,\tmes,\tmic)-n^{(f)}(|\q'_1|,\tmes,\tmic)\r]
~,
\eea
in which one interprets the multiplied coefficients in $n^{(f)}(|\q_1|,\tmes,\tmic)$ as the graviton absorption rate and the multiplied coefficients in $n^{(f)}(|\q'_1|,\tmes,\tmic)$ as the graviton emission rate. Also, we note that throughout the paper, we assume that $n^{(f)}$ is isotropic and is a function of energy $q^0=|\q|$. 
The oscillating terms like $e^{i(p^0_2+q^0_2-q'^0_2)\tmic}$ can be ignored using the secular approximation, as in the following. 
The secular approximation states that one can remove fast-oscillating terms in the interaction picture in which 
$p^0_2\neq q'^0_2-q^0_2 $. This approximation is feasible as far as the frequencies are well spaced in such a way
 that $|p^0_2+q^0_2-q'^0_2|^{-1} \ll \tmic \ll \tmes$. 
After confirming this approximation, we can eliminate the term $e^{-i(p^0_2+q^0_2-q'^0_2)\tmic}$
 by assuming $p^0_2+q^0_2-q'^0_2\approx 0$. As stated in previous sections, for a QBE describing a Markovian process, energy conservation occurs automatically, and no secular approximation is required,
 but for the GW damping effect, we have to use the secular approximation to establish energy conservation.
 
After calculating the dissipation term and inserting it into the QBE \eqref{boltzmanneq3}, we derive a full system of equations that describes the time evolution of the intensity and the polarization of GWs
\bea \label{boltzmanneq34}
(2\pi)^3\delta^3(0)2k^0\frac{d}{dt_{\textrm{mes}}}\rho^{\g}_{ij}(\k,\x,t_{\textrm{mes}})&=&-
\frac{\kappa^2}{8}\int_{0}^{\tmes}dt_{\textrm{mic}}\int d\q_1d\q'_1(2\pi)^6\delta^3(\k+\q_1-\q'_1)\delta^3(\q'_1-\q_1-\k )
\nonumber \\ & &\!\!\!\! \!\!\!\!\times\sum_{s_1,r_1,r'_1}q_1^{m_1}q_2^{m_2}
h^{s_1}_{m_1n_1}(\k) h^{\ast\, j}_{m_2n_2}(\k)\bar{u}_{r'_1}(\q'_1)\gamma^{n_1}u_{r_1}(\q_1)
 \bar{u}_{r_1}(\q_1)\gamma^{n_2}u_{r'_1}(\q'_1)
\nonumber \\ & \times &  
\rho^{\g}_{s_1i}(\k,\x,\tmes-\tmic)\l [n^{(f)}(|\q_1|,\tmes,\tmic)
\right.\nonumber \\ && \left.~~~~~~~~~~~~~~~~~~~~~~~~~~~~~~~~~~~
-n^{(f)}(|\q_1|+|\k|,\tmes,\tmic)\r]~,
\eea
where, in the last step, we have used the conservation of energy $|\q'_1|=|\q_1|+|\k|$.
The explicit time-dependence form of $n^{(f)}$
can be obtained from the environment dynamics. In fact, one must write the same QBE as \eqref{boltzmanneq3} for the environment and $n^{(f)}$.
 However, as it was discussed, the environment is large and its dynamics is fast
enough in such a way that its energy exchanged with the system will quickly dissipate away. Then, from the viewpoint of the
system, the state of the environment will appear to be almost constant all the time, and the environment is in thermal equilibrium.
The main difference with a conventional non-Markovian process in the physics of condense matter is that here, the dependence on $\tmic$ is due to the expansion of space-time.
It should also be noted that $\rho^{\g}_{s_1i}(\k,\x,\tmes-\tmic)$ is dependent on $\tmic$. However, the non-local kernel in the collision term arises also due to the dependence of $n^{(f)}$ on $\tmic$ and the expansion of space-time. The property of the dependence of $n^{(f)}$ on microscopic time has an important role in the non-Markovianity of the system's dynamics. As we will discuss below, $n^{(f)}$ depends on the expansion history of the expanding universe between $\tmes$ and $\tmic$.



\section{Damping of GWs by decoupled neutrinos }

It has been showed explicitly that the propagation of a gravitational wave in an environment containing decoupled relativistic neutrinos leads to the GW damping effect \cite{Weinberg:2003ur}. 
In fact, the environment induces a damping anisotropic stress tensor for the time evolution of the metric perturbation $h_{ij}$. The damping effect due to such an anisotropic stress tensor has been also predicated earlier \cite{Hawking:1966qi,Gayer:1979ff,Polnarev,Szekeres:1971ss,Weinberg:1972kfs}. 
By considering the contribution of free-streaming neutrinos to the anisotropic stress, an integro-differential equation for the GW propagation in the environment is obtained \cite{Weinberg:2003ur}. After numerically solving this equation and obtaining the amplitude of the gravitational wave, it is shown that the damping impact on the cosmological gravitational wave is not negligible.

The GW dynamics is given by 
\beq
\frac{d^2}{d\tmes^2}h_{ij}(\k,\tmes)+3H(\tmes)\frac{d}{d\tmes}h_{ij}(\k,\tmes)+\frac{k^2}{a(\tmes)}h_{ij}(\k,\tmes)= \kappa^2 \pi_{ij}(\k,\tmes)~,
\eeq
where $\pi_{ij}$ is the  tensor component of the matter anisotropic stress \cite{Weinberg:2003ur}, $H$ is the Hubble expansion rate, and $a(\tmes)$ is the scale factor given by the following line element in Friedmann-Lema\^itre-Robertson-Walker (FLRW) background,
\beq
ds^2=d\tmes^2-a^2(\tmes)\delta_{ij}dx^idx^j~.
\eeq
The mode dynamics is influenced by the Hubble parameter $H$ and by the anisotropic stress source term $\pi_{ij}$. It was shown that 
all short wavelength tensor modes reentering the horizon during the radiation dominated era, from the epoch of neutrino decoupling to the matter domination era,
are suppressed by a factor $A_0 \approx 0.8$ \cite{Weinberg:2003ur}. 

Here, we turn to calculate the term derived in \cite{Weinberg:2003ur} using the approach of open quantum systems and QBE. To this end, we assume 
that soft gravitons interact with a medium involving relativistic fermions with momentum $\q =|\q|(\sin\theta\cos\phi,\sin\theta\sin\phi,\cos\theta)$. 
The relativistic fermions are described by the following right-handed and left-handed helicity eigenstates \cite{Itzykson},
\beq
u_+(\hat{\q})=\frac{1}{\sqrt{2}}\left(\begin{array}{c} \cos(\theta/2) \\ \sin(\theta/2)e^{i\phi} \\ \cos(\theta/2) \\ \sin(\theta/2)e^{i\phi}  \end{array}\right)
~,~~~~~~~~~~~
u_-(\hat{\q})=\frac{1}{\sqrt{2}}\left(\begin{array}{c} -\sin(\theta/2)e^{-i\phi}  \\ \cos(\theta/2)\\ \sin(\theta/2) e^{-i\phi} \\ -\cos(\theta/2) \end{array}\right)~.
\eeq
Now, we further simplify \eqref{scattering3} by taking integration over $\q'_1$,
\bea  \label{collision1}
D_{ij}[\rho^{\g}(\k,\x,t_{\textrm{mes}})]&=&-\delta^3(0)
\frac{\kappa^2}{8}\int_{0}^{\tmes}dt_{\textrm{mic}}\int d^3q
\nonumber \\ & \times & \sum_{s_1,r_1,r'_1}q_1^{m_1}q_2^{m_2}
h^{s_1}_{m_1n_1}(\hat{\k}) h^{\ast\, j}_{m_2n_2}(\hat{\k})\bar{u}_{r'_1}(\hat{\q}+\hat{\k})\gamma^{n_1}u_{r_1}(\hat{\q})
 \bar{u}_{r_1}(\hat{\q})\gamma^{n_2}u_{r'_1}(\hat{\q}+\hat{\k})
\nonumber \\ & \times &  
\rho^{\g}_{s_1i}(\k,\x,\tmes-\tmic)\l [n^{(f)}(|\q|,\tmes,\tmic)-n^{(f)}(|\q|+|\k|,\tmes,\tmic)\r]
~,
\eea
where we have relabeled the momentum $\q_1$ as $\q$.
By plugging the helicity states $u_{\pm}$ into \eqref{collision1} and taking summation over both $r_1, r_2=+,-$, we get
\bea  \label{collision2}
D_{ij}[\rho^{\g}(\k,\x,t_{\textrm{mes}})]&=&-\delta^3(0)
\frac{\kappa^2}{4}\int_{0}^{\tmes}dt_{\textrm{mic}}\int d^3q  \sin^2\theta \l(\q\cdot \e^{(j)}\r)^2 
\nonumber \\ & \times & g_f
\l[n^{(f)}(|\q|,\tmes,\tmic)-n^{(f)}(|\q|+|\k|,\tmes,\tmic)\r]
\nonumber \\ & \times &  
\l[\cos2\phi\rho^{\g}_{1i}(\k,\x,\tmes-\tmic)+
\sin2\phi\rho^{\g}_{2i}(\k,\x,\tmes-\tmic)
\r]
~
\eea
where $g_f$ denotes the number of helicity states for fermions.
Therefore, using this result, the QBE \eqref{boltzmanneq34} becomes 
\bea \label{boltzmanneq3-3}
\frac{d}{dt_{\textrm{mes}}}\rho^{\g}_{ij}(\k,\x,t_{\textrm{mes}})&=&\frac{\kappa^2g_f}{8|\k|}\int_{0}^{\tmes}dt_{\textrm{mic}}\int \frac{d^3q}{(2\pi)^3}  \sin^2\theta \l(\q\cdot \e^{(j)}\r)^2 
|\k|\frac{\partial}{\partial |\q|} n^{(f)}(|\q|,\tmes,\tmic)
\nonumber \\ & \times &  
\l[\cos2\phi\,\rho^{\g}_{1i}(\k,\x,\tmes-\tmic)+
\sin2\phi\,\rho^{\g}_{2i}(\k,\x,\tmes-\tmic)
\r]~,
\eea
where, in the last step, we have expanded $n^{(f)}$ up to the first order in $|\k|$. 
We are interested in the time evolution of the intensity of GWs defined by $I^{(\textrm{g})}=\rho^{\g}_{11}+\rho^{\g}_{22}$. To evaluate $I^{(\textrm{g})}$, we assume that gravitons are propagating in $z$ direction with the following 
basis for direction and polarization vectors:
\bea
\k &=&(0,0,1)~,\\
\e^{(1)}&=&(1,0,0)~,\\
\e^{(2)}&=&(0,1,0)
~.
\eea 
The relativistic fermions are also described by an unpolarized Fermi-Dirac distribution
\beq
n^{(f)}(|\q|,\tmes,\tmic)= U_f(\tmes,\tmic)n_f(|\q|,\tmes)~, 
\eeq
with 
\beq
n^{(f)}(|\q|,\tmes)=\l [e^{\frac{|\q|}{T_f}}+1\r]^{-1}~,
\eeq
where $T_f$ is the temperature associated with ultra-relativistic fermions, and $U_f$ is the time evolution operator.
Now, inserting $n_f$ into the equation \eqref{boltzmanneq3-3}, using integration by parts and also integrating over $\phi$, we find the time evolution of the intensity as 
\bea \label{intensity0}
\dot{I}^{(\textrm{g})}(k,\x,\tmes)&=&-\frac{\kappa^2g_f}{8\pi^2}\int_{0}^{\tmes}dt_{\textrm{mic}}\int d|\q| d(\cos\theta) \sin^4\theta
 |\q|^3  \,U_f(\tmes,\tmic)
 \nonumber \\ & \times &  
\l[ e^{\frac{|\q|}{T_f}}+1\r]^{-1}
I^{\g}(k,\x,\tmes-\tmic)~,
\eea 
where dot denotes differentiation with respect to $\tmes$.
 Integrating over $|\q|$ gives
\bea \label{intensity0-1}
\dot{I}^{(\textrm{g})}(k,\x,\tmes)&=&-\frac{\kappa^2\bar{\rho}_{f} }{4}\int_{0}^{\tmes}dt_{\textrm{mic}}\int d(\cos\theta) \sin^4\theta\,
U_f(\tmes,\tmic)
I^{\g}(k,\x,\tmes-\tmic)~,
\eea 
where $\bar{\rho}_{f}$ is the total energy of fermions
per proper volume,
\beq
\bar{\rho}_f=g_f\frac{4\pi}{(2\pi)^3}\int d|\q| |\q|^3\l[ e^{\frac{|\q|}{T_f}}+1\r]^{-1}=\frac{7}{8}\frac{\pi^2}{30}g_f\,T^4_f~.
\eeq
The time evolution of $U_f(\tmes,\tmic)$ has been computed in \cite{Weinberg:2003ur} and is parameterized as
\beq 
U_f(\tmes,\tmic)=
e^{-i\int_{\tmic}^{\tmes}dt'\frac{|\mathbf{ k}|\mu}{a(t')}}~,
\eeq 
where $\mu=\hat{\q}\cdot\hat{\mathbf{ k}}=\cos\theta$ and $U_f$ satisfy the semigroup property as expected.
 Therefore, we get
\bea \label{intensity0-1}
\dot{I}^{(\textrm{g})}(k,\x,\tmes)=-\frac{\kappa^2}{4}\bar{\rho}_{f}\int_{0}^{\tmes}dt_{\textrm{mic}}\int d\mu (1-\mu^2)^2
\,e^{-i\int_{\tmic}^{\tmes}dt'\frac{|\mathbf{ K}|\mu}{a(t')}}
I^{\g}(k,\x,\tmes-\tmic)~,
\eea 
Now, using the integrating by parts technique to take integration over $|\k|$ and also taking the integration over $\mu$ we find
\bea \label{eqintensity1-0}
\dot{I}^{(\textrm{g})}(k,\x,\tmes)
= -
4\kappa^2\bar{\rho}_f\int_{0}^{\tmes}dt_{\textrm{mic}}\,\frac{j_2\l(s\r)}{s^2}
I^{(\textrm{g})}(k,\x,\tmes-\tmic)~,
\eea 
where $j_2(s)$ is the spherical Bessel function, and
\beq \label{s}
s=|\k|\int_{\tmic}^{\tmes}\frac{dt'}{a(t')}~.
\eeq
Equation \eqref{eqintensity1-0}, describing the macroscopic time evolution of the GW intensity equation, arises from the quantum master equation and involves a memory kernel. 
The memory effect is due to the time non-locality property of $I^{(\textrm{g})}(k,\x,\tmes-\tmic)$. Moreover, the expansion history of the FLRW 
universe will produce a memory effect as well. The memory kernel $j_2(s)$ is non-local in time due to the fact that it keeps memory about the starting point $\tmic$.  The appearance of $\tmic$ in the memory kerne implies that the dynamics contains a memory. Therefore, the equation \eqref{eqintensity1-0} is non-Markovian and keeps any memory about $\tmic$ (see \cite{Dariusz,Agarwal} for a discussion about non-Markovian conditions of an open quantum system).

The full time derivative of $I^{(\textrm{g})}$ can be expanded into partial derivatives as
\bea \label{eqintensity-full}
\frac{\partial}{\partial \tmes}I^{(\textrm{g})}(k,\x,\tmes)-Hk\frac{\partial}{\partial k}I^{(\textrm{g})}(k,\x,\tmes)+\frac{1}{a}\,\hat{\k}\cdot \bm{\nabla}I^{(\textrm{g})}(k,\x,\tmes)
&=&-
4\kappa^2\bar{\rho}_f(\tmes)\int_{0}^{\tmes}dt_{\textrm{mic}}\,\frac{j_2\l(s\r)}{s^2}
\nonumber \\ &\times&
I^{(\textrm{g})}(k,\x,\tmes-\tmic)~.
\eea 
The energy density of gravitational waves is given by \cite{Bartolo:2018igk}
\beq \label{intensity1}
\rho^{(\textrm{g})}(\x,\tmes)=\frac{1}{2}\left<\dot{h}_{ij}(\x,\tmes)\dot{h}^{ij}(\x,\tmes)\right>=\frac{1}{4} \int \frac{d^3 k}{(2\pi)^3}k^0
 I^{(\textrm{g})}(k,\x,\tmes)~.
\eeq 
Using this expression and after taking integration over $k$, one can rewrite the equation \eqref{eqintensity-full} in terms of energy density in the following form:
\bea \label{energydensity0}
\frac{\partial}{\partial \tmes}\rho^{(\textrm{g})}(\x,\tmes)+4H \rho^{(\textrm{g})}(\x,\tmes)+\frac{1}{a}\,\bm{\nabla}\cdot \bm{\mathcal{P}}^{(\textrm{g})}(\x,\tmes)
&=& -
4\kappa^2\bar{\rho}_f(\tmes)\int_{0}^{\tmes}dt_{\textrm{mic}}\,\frac{j_2\l(s\r)}{s^2}
\nonumber \\ &\times&
\rho^{(\textrm{g})}(\x,\tmes-\tmic)~,
\eea 
where
\beq \label{GWmflux}
\bm{\mathcal{P}}^{(\textrm{g})}(\x,\tmes)=\frac{1}{2}\left<\dot{h}_{ij}(\x,\tmes)\bm{\nabla}h^{ij}(\x,\tmes)\right>=\frac{1}{4}\int \frac{d^3k}{(2\pi)^3}\k I^{(\textrm{g})}(k,\x,\tmes)~,
\eeq
is the momentum out-flux carried away by GW. 

In order to compare our results with \cite{Weinberg:2003ur}, we must extract the equation of motion for tensor fluctuations $h_{ij}(\x,\tmes)$. To this end, we write the non-local in time intensity $I^{(\textrm{g})}(k,\x,\tmes-\tmic)$ in the following form
\beq  \label{intensity2}
\frac{1}{4}\int \frac{d^3 k}{(2\pi)^3}k^0 I^{(\textrm{g})}(k,\x,\tmes-\tmic)=\frac{1}{2}\left<\dot{h}_{ij}(\x,\tmes)\dot{h}^{ij}(\x,\tmes-\tmic)\right>~.
\eeq 
Substituting \eqref{intensity1} and \eqref{intensity2} in \eqref{eqintensity-full} and after some straightforward calculations, we get
\beq \label{GWeq0}
\ddot{h}_{ij}(\x,\tmes)+3H\dot{h}_{ij}(\x,\tmes)
-\frac{\nabla^2}{a^2(\tmes)}\,h_{ij}(\x,\tmes)=\kappa^2\pi_{ij}(\x,\tmes)~,
\eeq
where the anisotropic stress tensor $\pi_{ij}$ is given by 
\bea \label{stresstensor}
\pi_{ij}(\x,\tmes)&=& -2\bar{\rho}_f(\tmes)\int_{0}^{\tmes}dt_{\textrm{mic}}\,\frac{j_2\l(s\r)}{s^2}\dot{h}_{ij}(\x,\tmes-t_{\textrm{mic}})
~.
\eea 
To compare this result with \cite{Weinberg:2003ur}, it should be noted that for left-handed relativistic neutrinos, we have $g_f = 1$. 
Considering neutrino and antineutrino, the neutrino energy density is given by
\beq
\bar{\rho}_f=6\bar{\rho}_\nu=2\times\frac{7}{8}\frac{\pi^2}{30}\,T^4_\nu~.
\eeq
Therefore,
\bea \label{stresstensor}
\pi_{ij}(\x,\tmes)&=& -4\bar{\rho}_\nu(\tmes)\int_{0}^{\tmes}dt_{\textrm{mic}}\,\frac{j_2\l(s\r)}{s^2}\dot{h}_{ij}(\x,\tmes-t_{\textrm{mic}})
~.
\eea 
In the following, we shall instead work with equation \eqref{eqintensity-full} and look for an analytical solution that reflects the effect of the fermionic environment on the GW intensity.


\subsection{Calculation of GW damping in the radiation-dominated era }

Here, we will find an approximate analytic solution of the evolution equation \eqref{eqintensity-full} in the radiation dominated era. The equation \eqref{eqintensity-full} is an equation with memory effects, which has the form of an integro-differential equation that is non-local in time. At the early stage of radiation dominance, one can consider the short memory or equivalently the Markovian approximation. In this approximation, $\mathcal{I}^{(\textrm{g})}$ will be independent of $\tmic$. The reason for this approximation is as follows. In the limit $s\rightarrow 0$, we can write $j_2(s)/s^2\rightarrow 1/15$. We instead replace $j_2(s)/s^2$ with $\alpha/15$ where $\alpha$ is a fudge factor that then we match to the numerical solution. Therefore, for the modes that are superhorizon during the radiation-dominated era one can write 
\bea \label{eqintensity1}
\frac{\partial}{\partial \tmes}\mathcal{I}^{(\textrm{g})}(k, \K,\tmes) -Hk\frac{\partial}{\partial k}\mathcal{I}^{(\textrm{g})}(k,\K,\tmes)-\frac{i}{a(\tmes)}\,\hat{\k}\cdot \K\,\mathcal{I}^{(\textrm{g})}(k, \K,\tmes)
&=&-
\frac{8\alpha\kappa^2}{15}\bar{\rho}_\nu(\tmes)
\nonumber \\ && \!\!\!\!\!\!\!\!\!\!\!\!\!\!\!\!\!\!\!\!\!\!\!\!\!\!\!\!\!\!\!\!\!\!\! \!\!\!\!\!\!\!\!\!\!\!\!\!\!\!\!\!\!\!\!\!\!\!\!\!\!\!\! \times
\int_{0}^{\tmes}dt_{\textrm{mic}}
\mathcal{I}^{(\textrm{g})}(k, \K,\tmes-\tmic)~,
\eea 
where $\mathcal{I}^{(\textrm{g})}(k,\K,\tmes)$ is the Fourier transform of the intensity, and $\mathbf{ K}$ is the momentum conjugate to $\x$. The right-hand side of this equation is still dependent of $\tmic$. However, one can neglect the microscopic time in $\mathcal{I}^{(\textrm{g})}(k, \K,\tmes-\tmic)$ at the early stage of radiation dominance and  rewrite the right-hand side of \eqref{eqintensity1} in terms of 
$\mathcal{I}^{(\textrm{g})}(k, \K,\tmes)$ and $\tmes$ as in the following form:
\bea \label{eqintensity1-1}
\frac{\partial}{\partial \tmes}\mathcal{I}^{(\textrm{g})}(k,\K,\tmes)-Hk\frac{\partial}{\partial k}\mathcal{I}^{(\textrm{g})}(k,\K,\tmes)-\frac{i}{a(\tmes)}\,\hat{\k}\cdot \K\,\mathcal{I}^{(\textrm{g})}(k, \K,\tmes)
&=&-
\frac{8\alpha\kappa^2}{15}\,\bar{\rho}_\nu(\tmes)\tmes 
\nonumber \\&& \!\!\!\!\!\!\!\!\! \times
\mathcal{I}^{(\textrm{g})}(k, \K,\tmes)~.
\eea 
Using this expression, one can write
\beq \label{eqintensity2}
\frac{\partial}{\partial \tmes}\mathcal{I}^{(\textrm{g})}(k, \K,\tmes)-Hk\frac{\partial}{\partial k}\mathcal{I}^{(\textrm{g})}(k,\K,\tmes)-\frac{i}{a(\tmes)}\,\hat{\k}\cdot \K\,\mathcal{I}^{(\textrm{g})}(k, \K,\tmes)
= -
\frac{1}{\tau}\,
\mathcal{I}^{(\textrm{g})}(k, \K,\tmes)~,
\eeq
where $\tau$ is the damping time of GW,
\beq \label{tau1}
\frac{1}{\tau(\tmes)}
=
\frac{8\alpha\kappa^2}{15}\,\bar{\rho}_\nu(\tmes)\tmes~.
\eeq

It has been shown that in the early stage of radiation dominance, the effect of damping time on the tensor modes that are superhorizon is not negligible. To illustrate this for the intensity, we first write the scale factor 
$a(t)$ after neutrino decoupling as 
\beq
a(\tmes)\simeq \Omega^{1/4}_R(2H_0\,\tmes)^{1/2}~,
\eeq
where $H_0$ is the Hubble constant and $\Omega_R$ is the radiation energy fraction. The Hubble parameter is also given by
\beq
H(\tmes)=\frac{1}{2\tmes}~.
\eeq
In this era, the neutrino energy density is given by
\beq
\kappa^2\bar{\rho}_\nu(\tmes)\simeq 6 f_\nu H^2~,
\eeq
where, for three neutrino species, $f_\nu =\Omega_\nu/\Omega_R\simeq 0.4$. Therefore, 
\beq \label{tau2}
\frac{1}{\tau(\tmes)}
=
\frac{16\alpha}{5}\frac{f_\nu}{\tmes}~.
\eeq
When the effect of absorption can be ignored, i.e., the condition $H \tau\gg 1$ is met, we recover the ordinary Liouville equation for the gravitational wave.
In this limit, using \eqref{energydensity0}, we can show that the energy density of GW behaves as $\rho^{(\textrm{g})}(\x,\tmes)\propto a^{-4}(\tmes)$ times an oscillatory factor that is due to the gradient term on the left-hand side of \eqref{energydensity0}. From the expression \eqref{intensity1} and in a comoving frame, one would expect a similar scaling for the intensity as $I^{(\textrm{g})}(k,\x,\tmes)\propto a^{-4}(\tmes)$. Now, to obtain such a result for $I^{(\textrm{g})}(k,\x,\tmes)$ from equation \eqref{eqintensity-full} one must assume $I^{(\textrm{g})}(k,\x,\tmes)\propto k^{-4}$. 
Given this scaling behavior (which implies  that the anisotropic stress tensor of neutrinos just affects the amplitude of GWs, but they do not produce a change in frequency), the evolution equation \eqref{eqintensity2} becomes
\bea \label{eqintensity4}
\frac{\partial}{\partial \tmes}\mathcal{I}^{(\textrm{g})}(k, \K,\tmes)+\l(\frac{1}{2}+\frac{4\alpha f_\nu}{5} \r)\frac{4}{\tmes}\mathcal{I}^{(\textrm{g})}(k, \K,\tmes)-i\,\frac{\mu' \widetilde{K}}{2\sqrt{\tmes}}\,\mathcal{I}^{(\textrm{g})}(k, \K,\tmes)
=0~.
\eea
where $\mu'=\hat{\K}\cdot\hat{\k}$ and
\beq
\widetilde{K}=\frac{2\sqrt{H_0}\Omega_M}{ \Omega^{3/4}_R}\frac{K}{K_{\textrm{eq}}}~,
\eeq
with $K_{\textrm{eq}}$ denoting the wave number of the mode that reenters the horizon at matter-radiation equality. The analytical solution of this equation is in the following form
\beq \label{intensityK}
\mathcal{I}^{(\textrm{g})}(k, \K,\tmes)=
\mathcal{I}^{(\textrm{g})}(k, \K,t_{\textrm{end}})\l(\sqrt{\frac{t_{\textrm{end}}}{\tmes}}\r)^{4(1+\delta)}e^{i\widetilde{K}\mu'\l(\sqrt{\tmes}-\sqrt{t_{\textrm{end}}}\r)}
~,
\eeq
where $\delta=8\alpha f_\nu/5$, and $\mathcal{I}^{(\textrm{g})}(k,\K,t_{\textrm{end}})$ is the GW intensity at the end of inflation. Taking the Fourier transform of \eqref{intensityK} yields
 \beq \label{intensityx}
I^{(\textrm{g})}(k, \x_0,\tmes)=\l(\sqrt{\frac{t_{\textrm{end}}}{\tmes}}\r)^{4+32f_\nu/5}\int \frac{d^3K}{(2\pi)^3}\mathcal{I}^{(\textrm{g})}(k, \K,t_{\textrm{end}})e^{i\widetilde{K}\mu'\l(\sqrt{\tmes}-\sqrt{t_{\textrm{end}}}\r)}e^{i\K\cdot\x_0}
~,
\eeq
where $\x_0$ is the observer location. Now, using the following general decomposition
 \beq
e^{i\mu' K x}=
\sum_{\ell=0}^{\infty}i^{\ell}(2\ell+1)j_{\ell}\l(Kx)\r)P_{\ell}(\mu')~,
\eeq
and the expansion
\beq
\mathcal{I}^{(\textrm{g})}(k,\K,t_{\textrm{end}})=\mathcal{I}^{(\textrm{g})}(k,K,t_{\textrm{end}})\sum_{LM}c^I_{LM}Y_L^M
(\hat{\K})~,
\eeq
 where $j_\ell(x)$ is the spherical
Bessel function, $P_\ell(\mu)$ is the Legendre polynomials, and $Y_L^M
(\hat{\K})$ is the spherical harmonics, we find at $\x_0=0$,
 \bea \label{intensityx1}
I^{(\textrm{g})}(k,\x_0=0,\tmes)&=&\l(\sqrt{\frac{t_{\textrm{end}}}{\tmes}}\r)^{4(1+\delta)}
\sum_{LM\ell}
\int \frac{d^3K}{(2\pi)^3}\mathcal{I}^{(\textrm{g})}(k,K,t_{\textrm{end}})
c^I_{LM}Y_L^M
(\hat{\K})
i^{\ell}(2\ell+1)
\nonumber \\ &\times&
j_\ell\l(\widetilde{K}\l(\sqrt{\tmes}-\sqrt{t_{\textrm{end}}}\r)\r)P_\ell(\mu')
~.
\eea
Taking the monopole term $L=M=0$, we get
 \bea \label{intensityx1}
I^{(\textrm{g})}(k,\x_0=0,\tmes)&=&\l(\sqrt{\frac{t_{\textrm{end}}}{\tmes}}\r)^{4(1+\delta)}
\frac{c^I_{00}}{4\pi^{5/2}}
\int dKK^2\mathcal{I}^{(\textrm{g})}(k,K,t_{\textrm{end}})
\nonumber \\ &\times&
j_0\l(\widetilde{K}\l(\sqrt{\tmes}-\sqrt{t_{\textrm{end}}}\r)\r)~.
\eea
At the early stages of the radiation-dominated epoch, the interaction with the fermion bath decreases the amplitude of the GW intensity by a factor $\delta\sim 0.64 \alpha$.

At the end of this section, we will estimate the factor $\alpha$ by comparing our results with \cite{Weinberg:2003ur} and \cite{Dicus:2005rh}. We first
 show that the same suppression factor $ \delta\sim 0.48\alpha$ could be also obtained from Eq.\eqref{GWeq0}. 
Considering the approximation $j_2(s)/s^2\rightarrow \alpha/15$ in the stress tensor \eqref{stresstensor}, one finds 
\beq \label{GWeq2}
\ddot{h}_{ij}(\x,\tmes)+\l(\frac{3}{2}+\frac{8\alpha f_\nu}{5} \r)\frac{1}{\tmes}\dot{h}_{ij}(\x,\tmes)
-\frac{\nabla^2}{a^2(\tmes)}\,h_{ij}(\x,\tmes)=0~.
\eeq
We write $h_{ij}(u)=h_{ij}(u_0)\chi(u)$, where in a radiation-dominated background, $u=2K\sqrt{\tmes}$. 
In the absence of the anisotropic stress, the solution of this equation becomes $\chi(u)=j_0(u)$ where $j_0$ is spherical Bessel function. In the presence, of the anisotropic stress the equation \eqref{GWeq2} can be solved approximately with the overall scaling $h_{ij}(\x,\tmes)\propto (\sqrt{\tmes})^{-1-2\delta} $ times an oscillating factor coming from the Laplacian term. Therefore, using \eqref{intensity1}, we find the time behavior of $\rho^{(\textrm{g})}(\x,\tmes)$ as
\beq \label{rhoij1}
\rho^{(\textrm{g})}(\x,\tmes)\propto a^2(\tmes)\dot{h}_{ij}(\x,\tmes)\dot{h}_{ij}(\x,\tmes)\propto \l(\frac{1}{\sqrt{\tmes}}\r)^{4(1+\delta)}~,
\eeq
which is the same as the scaling behavior that we found using \eqref{eqintensity4}. 

\begin{figure}
   \centering
     \includegraphics[width=6in,height=3in]{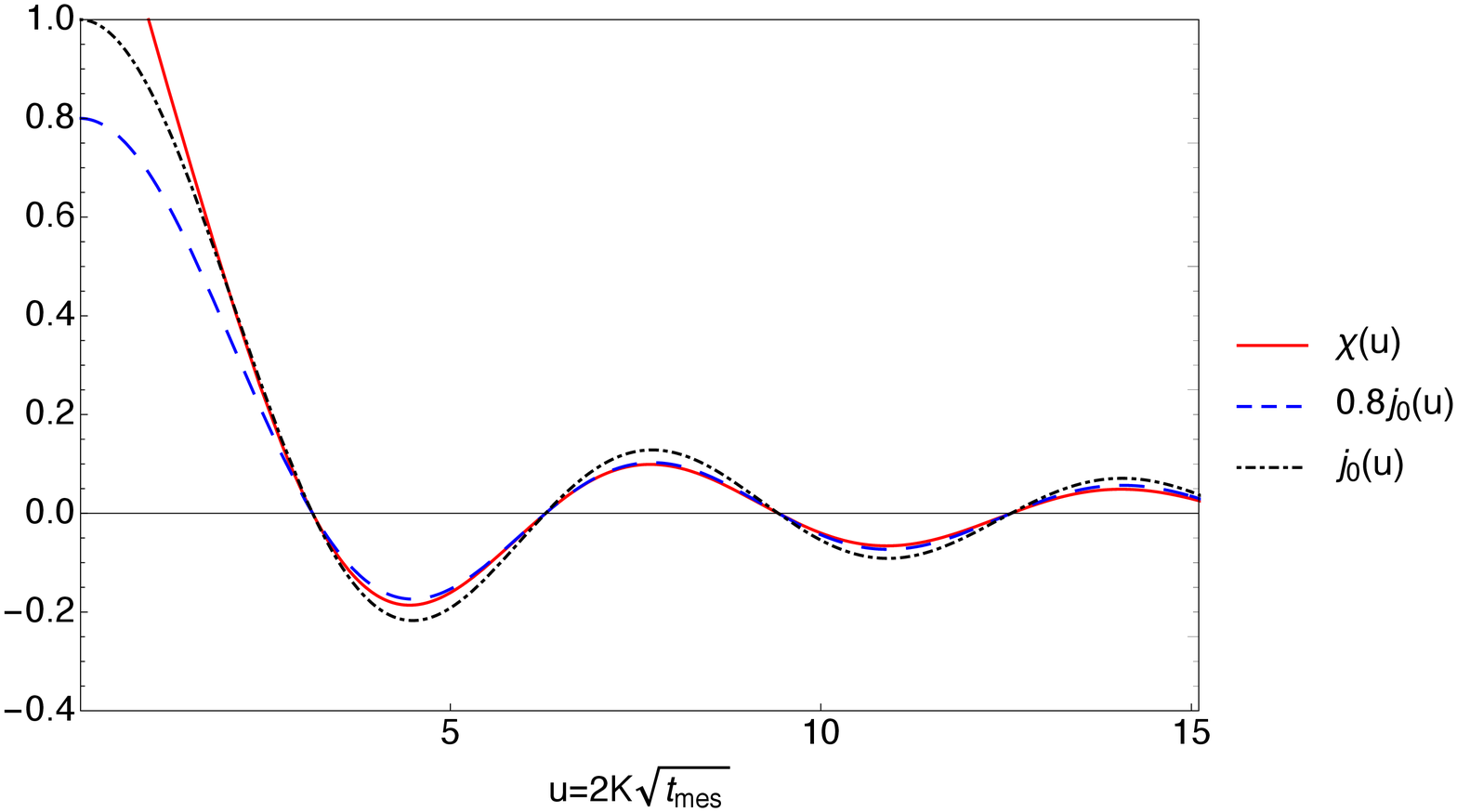}
        \caption{The solid-red curve shows the analytical solution Eq. \eqref{chi1} with $\alpha=2/7$ compared to the $f_\nu = 0$ solution $j_0(u)$ (dot-dashed, black) and the
asymptotic solution $0.8j_0(u)$ (dashed-blue). Here, $u=2K\sqrt{\tmes}$.}\label{plotGW}
    \end{figure}

The exact solution of Eq. \eqref{GWeq2} is also given as in the following form:
\beq \label{chi1}
\chi(u)=\frac{1}{2}\pi\chi(u_0)u_0^{2\delta+3/2}u^{-2\delta-1/2}\left(J_{2\delta+3/2}(u_0)Y_{2\delta+1/2}(u)-Y_{2\delta+3/2}(u_0)I_{2\delta+1/2}(u)\right)~,
\eeq
where $J_\alpha(u)$ and $Y_\alpha(u)$ are Bessel functions of the first and the second kind, respectively, and $\chi(u_0)$ is fixed by performing matching with the initial condition at $u_0$.
For $u>1$ the solution \eqref{chi1} can be approximated as follows:
\beq \label{chi1-2}
\chi(u)=\chi(u_0)\frac{\Gamma(2\delta+3/2)}{\pi^{1/2}}\left(\frac{2}{u}\right)^{2\delta+1}\sin(u)~.
\eeq
In Fig. \eqref{plotGW}, we have illustrated the analytical solution \eqref{chi1-2} for $\alpha=2/7$. We have also compared $\chi(u)$ with the solution $j_0(u)$ and the asymptotic solution $A j_0(u)$ with the suppression factor $A=0.8$ suggested in \cite{Weinberg:2003ur}. 
The $\chi(u)$ departs from the the $\delta=0$ solution $j_0(u)$ and approaches the asymptotic solution
$0.8 j_0(u)$ around $u=3$ as illustrated in Fig.  \eqref{plotGW}.

Now that we have fixed the parameter $\alpha$, the suppression factor of the GW intensity is obtained as $\delta=0.183$. It is worth mentioning that the SGWB also passes through the matter-dominated era until observed at present time. According to an argument given by \cite{Weinberg:2003ur},
we expect that \eqref{chi1-2} serves as an initial
condition for the subsequent evolution of the gravitational wave amplitude during the matter-dominated era. 
 Therefore, one should still expect the same change in the amplitude during the matter-dominated era.

\subsection{ Influence on the polarization of GW}

In this part, we provide the system of differential equations given by QBE that describe the time evolution of the GW's Stokes parameters in the radiation and matter-dominated epochs. 
To this end, we first define the vector $P$ in terms of the gravitational Stokes parameters as the form,
\beq
P^{(\textrm{g})},=\l(\mathcal{I}^{(\textrm{g})},\mathcal{Q}^{(\textrm{g})},\mathcal{U}^{(\textrm{g})},i\mathcal{V}^{(\textrm{g})}\r)~.
\eeq
Using this vector and assuming the scaling behavior as $P^{(\textrm{g})}(k,\K,\tmes)\propto k^{-4}$, one can represent the evolution equations \eqref{boltzmanneq3-3} in the following form:
\bea \label{pol}
\frac{\partial}{\partial \tmes}P^{(\textrm{g})}(k,\K,\tmes)+4HP^{(\textrm{g})}(k,\K,\tmes)-\frac{i}{a(\tmes)}\,\mu' KP^{(\textrm{g})}(k,\K,\tmes)
&=&-
8\kappa^2\bar{\rho}_\nu(\tmes)
\nonumber \\ && \!\!\!\!\!\!\!\!\!\!\!\!\!\!\!\!\!\!\!\!\!\!\!\!\!\!\!\!\!\!\!\!\!\!\!\!\!\!\!\!\!\!\!\!\!\!\!\!\!\!\!\!\!\!\!\!\!\!\!\!\!\!\!\!\!\!\!\!\!\!\!\!\!\!\!\!\!\!\!\!\!\!\!\!\!\!\!\!\!\!\!\!\!\!\!
\times\int_{0}^{\tmes}dt_{\textrm{mic}}\,\frac{j_2\l(s\r)}{s^2}
M\,P^{(\textrm{g})}(k,\K,\tmes,\tmic)~,
\eea
where 
\beq
M=\left(\begin{array}{cccc}1 & 0 & 0 & 0 \\0 & 1 & 0 & 0 \\0 & 0 & 1 & 0 \\0 & 0 & 0 & -1\end{array}\right)~,
\eeq
is the symmetric diagonal matrix. Therefore, we have a system of first-order time non-local differential equations that can be numerically integrated in an expanding universe background. Using equation \eqref{pol} and the same method as in the previous section, we can calculate the $V$ parameter during radiation dominance as follows:
\bea
V^{(\textrm{g})}(k,\x_0=0,\tmes)&=&
\l(\sqrt{\frac{t_{\textrm{end}}}{\tmes}}\r)^{4(1-\delta)}
\sum_{LM\ell}
\int \frac{d^3K}{(2\pi)^3}\mathcal{V}^{(\textrm{g})}(k,K,t_{\textrm{end}})
c^V_{LM}Y_L^M
(\hat{\K})
i^{\ell}(2\ell+1)
\nonumber \\ &\times&
j_\ell\l(\widetilde{K}\l(\sqrt{\tmes}-\sqrt{t_{\textrm{end}}}\r)\r)P_\ell(\mu')
~,
\eea
where $\mathcal{V}^{(\textrm{g})}(K,t_{\textrm{end}})$ is the V-mode parameter at the end of inflation. Very interestingly, the source term associated with the V-mode polarization of GWs changes sign. 
Therefore, contrary to the damping effect in the intensity and the linear polarization, the V-mode polarization is amplified by interaction with the medium containing decoupled relativistic fermions. The interaction causes $\rho^{\g}_{ij}\rightarrow \rho^{\g}_{ji}$, which, in turn, changes the sign of the $V$ parameter. The
 parity transformation of the interaction Hamiltonian 
 is a clue to identify the sign change of $V$. As it was shown, the interaction Hamiltonian is not invariant under parity transformation and interchanges the right- and left- handed circular polarization modes. Therefore, one can write
 \beq
 V=V_R-V_L  \xrightarrow{\text{P}}  V_L-V_R=-V~,
 \eeq
where $V_L$ and $V_R$ denote the left- and right- handed V-modes. 

\section{Conclusions}

We have discussed that QBE is a powerful and commonly used tool in the investigation of open quantum systems. In the conventional form of the QBE, the Born-Markov approximation is employed. The Markovian QBE is a time-local equation in which one ignores all memory effects. 
Recent studies show a wide range of applications of the Markovian QBE in the CMB, neutrino physics, and GWs. 
In this work, we consider memory effects beyond the Markovian approximation and generalize the QBE so that it can be used to describe the non-Markovian processes.  
Moreover, the conventional form of the Markovian QBE is only applicable to reversible processes. We discussed that for such processes the effective interaction Hamiltonian describing the process at the microscopic scales is invariant under time reversal operation. However, the effective interaction Hamiltonian associated with an irreversible process such as absorption transforms to the effective interaction Hamiltonian of the inverse process under time reversal transformation. 
We have extended the QBE to a new form that can also describe the irreversible phenomena. 
Among the many possible applications, in this work, we have used this new equation to explain the phenomenon of gravitational-waves damping during their propagation in an environment consisting of decoupled relativistic fermions using this approach. 
Using the non-Markovian QBE we obtained an integro-differential equation describing the irreversible dynamics of the reduced graviton system.
 Here, we computed the time evolution equation of the GW intensity, although we have shown that this equation is quite consistent with the equation previously calculated to explain the damping of the GWs due to free streaming neutrinos in the early universe. 

The non-Markovian equations are usually difficult to treat analytically. It would be very interesting to investigate the resulting non-Markovian equations for the density matrix of GWs numerically; however, it is left for future works. 
We instead considered the limit of the early stage of the radiation-dominated era during which, one can drastically reduce the complexity due to memory effects. 
The analytical solution for the GW intensity showed that interaction with the ultra-relativistic fermion environment gives rise to a damping effect that induces a suppression in the intensity by a factor $\delta=0.64\alpha$ more than the condition that there is no contact with the environment. 

We then fixed the parameter $\alpha$ in such a way that we first obtained an analytical solution for \eqref{chi1}, and by comparing our result with the asymptotic solution previously obtained by \cite{Weinberg:2003ur}, we found $\alpha=2/7$. Therefore, the suppression factor during radiation-dominated era was found as $\delta=0.183$.  

Additionally, another new aspect of our study, we showed the propagation of a circularly polarized SGWB in an environment of ultra-relativistic fermions that cause an enhancement to the Stokes parameter $V$ in contrast with the intensity and the linear polarization. This is because the interaction with the environment changes the handedness of the circularly polarized GWs that in turn causes $V\rightarrow -V$. This makes an amplification by a factor $\delta=0.64\alpha=0.183$ for the V-mode polarization. However, note 
that the polarized SGWB also passes through the matter-dominated era until observed at present time. The suppression obtained during the radiation dominated era provides the initial condition for the subsequent evolution during the matter-dominated era.

We can also make a general statement about the amplification of the V-mode polarization due to the coupling of GWs with a background matter field. 
In general, the linear interaction between the GWs and background matter is characterized through $\kappa\, h_{\mu\nu}T^{(M)\,\mu\nu}$, in which, $T^{(M)\,\mu\nu}$ describes the energy-momentum tensor of matter fields such as photon or dark matter. Inserting this interaction term into the QBE, one can verify that the absorption of a flux of circularly polarized GWs by a background of matter fields will amplify the V-mode polarization if $T^{(M)\,\mu\nu}$ is invariant under the parity transformation.


\begin{acknowledgments}
		
M.Z. acknowledges financial support by the University of Padova under the MSCA Seal of Excellence @UniPD programme. A.R. acknowledges funding from Italian Ministry of Education, University and Research (MIUR) through the `Dipartimenti di eccellenza' project Science of the Universe. N.B., D.B., and S.M. acknowledge partial financial support by ASI Grant No. 2016-24-H.0 and 2016-24-H.1-2018. M.Z. would like to thank S. Azaele, F. Baldovin, M. Saeedian, and F. Shahbazi for several enlightening and stimulating discussions. 
		
\end{acknowledgments}


\appendix

\section{Calculation of expectation values}

\subsection{Equal-time operators}

In the equilibrium condition, the creation and annihilation operators satisfy equal-time commutation relations
\beq \label{commtgr}
\l[a_s(p,t),a^\dag_{s'}(p',t)\r]=(2\pi)^32p^0\delta^3(\p-\p')\delta_{ss'}~, 
\eeq
and
\beq \label{commtf}
\l\{b_r(q,t),b^\dag_{r'}(q',t)\r\}=(2\pi)^3\delta^3(\q-\q')\delta_{rr'}~. 
\eeq
Using the above relations, one can calculate the connected equal-time expectation value of operators that is proportional to the density matrix. Before, we would like to emphasize that the correct expression for the expectation values containing all the information are those that involve connected pieces. The connected expectation value is defined as follows:
\beq
\l<a^{\dag }_1a_2\r>=\l<a^{\dag }_1\r>\l<a_2\r>+\l<a^{\dag }_1a_2\r>_\textrm{c}~.
\eeq
 With this in mind, we begin to calculate the equal-time expectation values for gravitons and ultra-relativistic fermions.
  First, we calculate the so called two-point equal-time expectation values. In general, the expectation value for gravitons is given by \cite{Bartolo:2018igk}
  \beq
  \l<a^{\dag }_{m}(p',t)a_{n}(p,t)\r>_\textrm{c}=\textrm{tr}\l[\hat{\rho}^{(\textrm{g})}a^{\dag }_{m}(p',t)a_{n}(p,t)\r]=
  \int \frac{d^3\p_1}{(2\pi)^3}\l<\p_1\r|\hat{\rho}^{(\textrm{g})}a^{\dag }_{m}(p',t)a_{n}(p,t)\l|\p_1\r>~.
  \eeq
  Now, using the graviton density operator \eqref{rhohatg}, we have 
  \beq
  \l<a^{\dag }_{m}(\p',t)a_{n}(\p,t)\r>_\textrm{c}= \int \frac{d^3\p_1}{(2\pi)^3}
  \int \frac{d^3\p_2}{(2\pi)^3} \rho^{(\textrm{g})}_{ij}(\p_2,t) \l<\p_1\r|a^{\dag}_i(\p_2,t)a_j(\p_2,t)a^{\dag }_{m}(\p',t)a_{n}(\p,t)\l|\p_1\r>~,
  \eeq
  where $\l|\p_1\r>$ is the one-particle graviton state with momentum $\p_1$ that is given by
  \beq
  \l|\p_1\r>=\frac{1}{2p^0_1}a^\dag_s(\p_1,t)\l|0\r>~.
  \eeq
  Therefore, using the commutation relation \eqref{commtgr}, we have
  \bea
  \l<a^{\dag }_{m}(\p',t)a_{n}(\p,t)\r>_\textrm{c}&= & \int \frac{d^3\p_1}{(2\pi)^3}
  \int \frac{d^3\p_2}{(2\pi)^3} \frac{1}{4(p^0_1)^2}\rho^{(\textrm{g})}_{ij}(\p_2,t)
  (2\pi)^9 8\omega_{\p_1}\omega_{\p_2} \omega_{\p}\delta_{li}\delta_{jm}\delta_{nl}
  \nonumber \\ && ~~~~~~~\times
  \delta^3(\p_1-\p_2)  \delta^3(\p_2-\p')  \delta^3(\p-\p_1)
  \nonumber \\ &=& 
  2p^0 (2\pi)^3\delta(\p-\p')\rho^{(\textrm{g})}_{nm}(\p,t)~.
  \eea
In the same way and using the anti-commutation relation \eqref{commtf}, we find
\beq 
\l<b^{\dag }_{m}(q',t)b_{n}(q,t)\r>_\textrm{c}=(2\pi)^3\delta^3(\q-\q')\rho^{(f)}_{nm}(\q,t)
~. 
\eeq

\begin{figure}
   \centering
     \includegraphics[width=4.5in,height=2in]{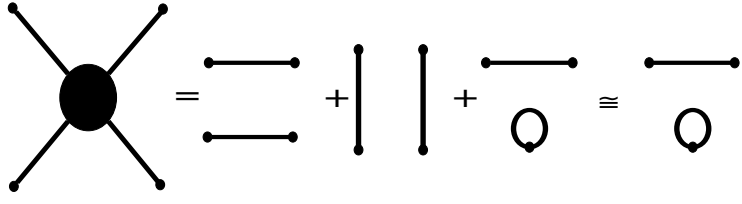}\label{axion22}
        \caption{Diagrammatic representation of the expectation values of four operators.}
    \end{figure}

The four-point equal-time expectation values are calculated as follows \cite{Kosowsky:1994cy,Danielewicz:1982kk}:
\bea 
\l<a^{\dag }_{s'_1}(\p'_1,t)a_{s_1}(\p_1,t)a^{\dag }_{s'_2}(\p'_2,t)a_{s_2}(\p_2,t)\r>_\textrm{c}&=&
4p^0_1p^0_2(2\pi)^6\delta^3(\p_1-\p'_1)\delta^3(\p_2-\p'_2)\rho^{\g}_{s_1s'_1}(\p_1,t)
\nonumber \\ &\times&
\rho^{\g}_{s_2s'_2}(\p_2,t)+
4p^0_1p^0_2(2\pi)^6\delta^3(\p_1-\p'_2)\delta^3(\p_2-\p'_1)
\nonumber \\ &\times&
\rho^{\g}_{s_2s'_1}(\p_2,t)\rho^{\g}_{s_1s'_2}(\p_1,t)+
2p^0_2(2\pi)^3\delta^3(\p_2-\p'_1)
  \nonumber \\ &\times&  \rho^{\g}_{s_2s'_1}(\p_2,t)
\l<\l [a_{s_1}(\p_1,t),a^{\dag }_{s'_2}(\p'_2,t)\r]\r>~,
\eea 
where the subscript ``c" denotes the sum of all diagrams connected to the external lines. The expectation value of the commutator is calculated as in the following, we write
\bea
\l<\l [a_{s}(\p,t),a^{\dag }_{s'}(\p',t)\r]\r>&=&
 \int \frac{d^3\p_1}{(2\pi)^3}
  \int \frac{d^3\p_2}{(2\pi)^3} (2\pi)^32 p^0\delta^3(\p-\p')\delta_{ss'}\rho^{(\textrm{g})}_{ij}(\p_2,t) \l<\p_1\r|a^{\dag}_i(\p_2,t)a_j(\p_2,t)\l|\p_1\r>
  \nonumber \\ &=&
  (2\pi)^32 p^0\delta^3(\p-\p')\delta_{ss'} N^{(\textrm{g})}
    \nonumber \\ &=&
    (2\pi)^32 p^0\delta^3(\p-\p')\delta_{ss'} ~,
\eea
where $N^{(\textrm{g})}$ is the number of gravitons,
\beq
N^{(\textrm{g})}=(2\pi)^3\delta^3(0)
  \int \frac{d^3\p_1}{(2\pi)^3} \rho^{(\textrm{g})}_{ii}(\p_1,t)~,
\eeq
and is assumed to be equal to 1. Therefore, the expected value of four operators is obtained in a similar way to the Wick's theorem as follows:
\bea
\l<a^{\dag }_{s'_1}(\p'_1,t)a_{s_1}(\p_1,t)a^{\dag }_{s'_2}(\p'_2,t)a_{s_2}(\p_2,t)\r>_\textrm{c}&=&
4p^0_1p^0_2(2\pi)^6\delta^3(\p_1-\p'_1)\delta^3(\p_2-\p'_2)\rho^{\g}_{s_1s'_1}(\p_1,t)
\nonumber \\ &\times&
\rho^{\g}_{s_2s'_2}(\p_2,t)+
4p^0_1p^0_2(2\pi)^6\delta^3(\p_1-\p'_2)\delta^3(\p_2-\p'_1)
\nonumber \\ &\times&
\rho^{\g}_{s_2s'_1}(\p_2,t)\rho^{\g}_{s_1s'_2}(\p_1,t)
+
4p^0_1p^0_2(2\pi)^6\delta^3(\p_2-\p'_1)
\nonumber \\ &\times&
\delta^3(\p_1-\p'_2)\delta_{s_1s'_2}\rho^{\g}_{s_2s'_1}(\p_2,t)
\nonumber \\ &\simeq & 
4p^0_1p^0_2(2\pi)^6\delta^3(\p_2-\p'_1)\delta^3(\p_1-\p'_2)
\nonumber \\ &\times&
\delta_{s_1s'_2}\rho^{\g}_{s_2s'_1}(\p_2,t)~,
\eea
where, in the last line, we have kept the linear term in terms of $\rho^{\g}$. In Fig. 2, we have diagrammatically shown this expression. 
In the same manner, we have
\bea \label{eq4pointfermion}
\l< b^{\dag }_{r'_1}(\q'_1,t)b_{r_1}(\q_1,t)b^{\dag }_{r'_2}(\q'_2,t)b_{r_2}(\q_2,t)\r>_\textrm{c}&=&
(2\pi)^6\delta^3(\q_1-\q'_1)\delta^3(\q_2-\q'_2)
\rho^{(f)}_{r_1r'_1}(\q_1,t)\rho^{(f)}_{r_2r'_2}(\q_2,t)
\nonumber \\ &-&
(2\pi)^6\delta^3(\q_1-\q'_2)\delta^3(\q_2-\q'_1)
\rho^{(f)}_{r_1r'_2}(\q_2,t)\rho^{(f)}_{r_2r'_1}(\q_1,t)
\nonumber \\ &+&
(2\pi)^3\delta^3(\q_2-\q'_1)\rho^{(f)}_{r_2r'_1}(\q_2,t)
\l<\l\{b_{r_1}(\q_1,t),b^{\dag }_{r'_2}(\q'_2,t)\r\}\r>
\nonumber \\ &\simeq & 
(2\pi)^3\delta^3(\q_2-\q'_1)\rho^{(f)}_{r_2r'_1}(\q_2,t)\l<\l\{b_{r_1}(\q_1,t),b^{\dag }_{r'_2}(\q'_2,t)\r\}\r>
\nonumber \\ &= & 
(2\pi)^6\delta^3(\q_2-\q'_1)\delta^3(\q_1-\q'_2)
\delta_{r_1r'_2}\rho^{(f)}_{r_2r'_1}(\q_2,t)
~,
\eea 
where we have assumed that the number of fermions is equal to 1, and therefore, the expectation value of anti-commutation relation is given by
\beq
\l<\l\{b_{r}(\q',t),b^{\dag }_{r'}(\q'_2,t)\r\}\r>= (2\pi)^3\delta^3(\q-\q')\delta_{rr'}~. 
\eeq 
In all of the above expectation values, we can take $t$ to be the mesoscopic time $\tmes$.

\subsection{Unequal-time expectation values}

As was mentioned, it is convenient that in equilibrium quantum field theory, we define the operator algebra for the bosonic and the fermionic creation and
annihilation operators by the equal-time canonical commutation and anti-commutation relations. However, for out-of-equilibrium calculations, there are additional complications that do not appear in equilibrium condition. One of them is that in nonequilibrium conditions, we often have to apply time-dependent commutation or anti-commutation relations. In such condition, the Schwinger-Keldysh, in-in or closed-time-path formalism \cite{Schwinger:1960qe,Keldysh:1964ud} is applied to calculate the evolution and expectation values of observables that usually are in the Heisenberg picture. This formalism has been applied in cosmology, condensed matter problems, and in studying heavy ion collisions (for a detailed discussion, we refer to Refs. \cite{Rammer,Kamenevej,Danielewicz:1982kk,Bruus,Berges:2004yj}). In particular, it has been used for computing cosmological correlations during cosmological inflation as well as during preheating after inflation, for the dynamics of phase transitions in the early universe, and to study the dynamics of baryogenesis \cite{Boyanovsky:1997mq,Chen:2017ryl}. 

Here, we present a general discussion of the calculation of the time correlation functions. 
For a system at equilibrium conditions, such correlation functions depend only on the
time interval $t-t'$,
\beq
\l<A(t)B(t')\r>=C(t-t')~.
\eeq
For instance, we compute the non-equal-time expectation values for gravitons under the assumption that 
gravitons do not interact with the environment,
\bea
  \l<a^{\dag }_{m}(\p',t')a_{n}(\p,t)\r>_\textrm{c}&=& \int \frac{d^3\p_1}{(2\pi)^3}
  \int \frac{d^3\p_2}{(2\pi)^3} \rho^{(\textrm{g})}_{ij}(\p_2,\tau) \l<\p_1,\tau\r|a^{\dag}_i(\p_2,\tau)a_j(\p_2,\tau)a^{\dag }_{m}(\p',t')a_{n}(\p,t)\l|\p_1,\tau\r>
  \nonumber \\ &=& 
  \int \frac{d^3\p_1}{(2\pi)^3}
\frac{1}{2p^0_1} \rho^{(\textrm{g})}_{lj}(\p_1,\tau) \l<0\r|  \l[a_j(\p_1,\tau),a^{\dag }_{m}(\p',t')\r] \l[a_{n}(\p,t),a^{\dag }_{l}(\p_1,\tau)\r]\l|0\r>
 \nonumber \\ &=& 
 (2\pi)^3 2p'^0\delta^3(\p-\p')\rho^{(\textrm{g})}_{mn}(\p',\tau)C(t-t')~,
  \eea
  where $C(t)$ is time-dependent Green's function of creation and annihilation operators, defined as
  \beq
\l <0\r|  \l[a_m(\p_1,t_1),a^{\dag }_{n}(\p_2,t_2)\r]\l|0\r> = (2\pi)^3 2p^0_1\delta^3(\p_1-\p_2)\delta_{mn}C(t_1-t_2)~,
  \eeq
with the semigroup property that $C(t_1)C(t_2)=C(t_1+t_2)$ \cite{Kamenevej}. Therefore, we can write 
\bea
  \l<a^{\dag }_{m}(\p',t')a_{n}(\p,t)\r>_\textrm{c} =
2p^0(2\pi)^3\delta^3(\p-\p') \rho^{(\textrm{g})}_{nm}(\p,t-t')~.
  \eea
In the same way, for fermions we have
\beq
\l<b^{\dag }_{m}(q',t')b_{n}(q,t)\r>_\textrm{c}=(2\pi)^3\delta^3(\q-\q')\rho^{(f)}_{nm}(\q,t-t')~.
\eeq
We are interested in the four-point unequal-time expectation values. As it was assumed, the environment is in the equilibrium state. Therefore,
for the fermions, we can write
 \bea \label{noneq4pointfermion}
\l< b^{\dag }_{r'_1}(q'_1,t'_1)b_{r_1}(q_1,t_1)b^{\dag }_{r'_2}(q'_2,t'_2)b_{r_2}(q_2,t_2)\r>_\textrm{c}&=&
(2\pi)^6\delta^3(\q_1-\q'_1)\delta^3(\q_2-\q'_2)\rho^{(f)}_{r_1r'_1}(\q_1,t_1-t'_1)\
\nonumber \\ & \times&
\rho^{(f)}_{r_2r'_2}(\q_2,t_2-t'_2)-
(2\pi)^6\delta^3(\q_1-\q'_2)\delta^3(\q_2-\q'_1)
\nonumber \\ & \times&
\rho^{(f)}_{r_1r'_2}(\q_1,t'_2-t_1)\rho^{(f)}_{r_2r'_1}(\q_2,t_2-t'_1)+
(2\pi)^6\delta^3(\q_1-\q'_2)
\nonumber \\ & \times&
\delta^3(\q_2-\q'_1)\delta_{r_1r'_2}\rho^{(f)}_{r_2r'_1}(\q_2,t_2-t'_1)
\nonumber \\ &\simeq & 
(2\pi)^6\delta^3(\q_1-\q'_2)\delta^3(\q_2-\q'_1)\delta_{r_1r'_2}
\nonumber \\ & \times&
\rho^{(f)}_{r_2r'_1}(\q_2,t_2-t'_1)
~,
\eea 
where, in the last line, we have kept the linear term in terms of $\rho^{(f)}$. Based on what was discussed, the system of gravitons is in out-of-equilibrium conditions, and therefore, the semigroup condition no longer applies to it. Accordingly, we write the two-point unequal-time expectation value as in the following form:
\bea
 \l<a^{\dag }_{m}(\p',t')a_{n}(\p,t)\r>_\textrm{c} =
2p^0(2\pi)^3\delta^3(\p-\p') \rho^{(\textrm{g})}_{nm}(\p,t,t')~.
 \eea
The four-point unequal-time expectation value is represented in the following form: 
 \bea \label{noneq4pointgraviton}
\l<a^{\dag }_{s'_1}(p'_1,t'_1)a_{s_1}(p_1,t_1)a^{\dag }_{s'_2}(p'_2,t'_2)a_{s_2}(p_2,t_2)\r>_\textrm{c}&=&
4p^0_1p^0_2(2\pi)^6\delta^3(\p_1-\p'_1)\delta^3(\p_2-\p'_2)\rho^{\g}_{s_1s'_1}(\p_1,t_1,t'_1)
\nonumber \\ & \times&
\rho^{\g}_{s_2s'_2}(\p_2,t_2,t'_2)+
4p^0_1p^0_2(2\pi)^6\delta^3(\p_1-\p'_2)\delta^3(\p_2-\p'_1)
\nonumber \\ & \times&
\rho^{\g}_{s_2s'_1}(\p_2,t_2,t'_1)\rho^{\g}_{s_1s'_2}(\p_1,t'_2,t_1)
+
4p^0_1p^0_2(2\pi)^6\delta^3(\p_1-\p'_2)
\nonumber \\ & \times&
\delta^3(\p_2-\p'_1)\delta_{s_1s'_2}\rho^{\g}_{s_2s'_1}(\p_2,t_2,t'_1)
\nonumber \\ &\simeq & 
4p^0_1p^0_2(2\pi)^6\delta^3(\p_1-\p'_2)\delta^3(\p_2-\p'_1)\delta_{s_1s'_2}
\nonumber \\ &\times & 
\rho^{\g}_{s_2s'_1}(\p_2,t_2,t'_1)~.
\eea 
In the same manner, we can calculate the following expressions:
\bea 
\l<a_{s_1}(p_1,t_1)a^{\dag}_{s_2}(p_2,t_2)a^{\dag}_i(k,t)a_j(k,t)\r>_\textrm{c}&\simeq&
(2\pi)^6 4p_1^0k^0\delta^{(3)}(\p_1-\p_2)\delta^3(0)\delta_{s_1s_2}\rho_{ji}(\k,t)
\nonumber \\ &+&  
(2\pi)^6 4p_1^0k^0\delta^{(3)}(\p_1-\k)\delta^3(\k-\p_2)
\nonumber \\ &\times & 
\delta_{s_1i}\rho_{js_2}(\k,t,t_2)~,
\eea 
and
\bea 
\l<a_{s_1}(p_1,t_1)a^{\dag}_i(k,t)a_j(k,t)
a^{\dag}_{s_2}(p_2,t_2)\r>_\textrm{c}&\simeq&
(2\pi)^6 4k^0p_1^0\delta^{(3)}(\mathbf{p}_1-\mathbf{p}_2)\delta^{(3)}(0)\delta_{s_1s_2}\rho_{ji}(\k,t)
\nonumber \\ &+& 
(2\pi)^6 4k^0p_1^0\delta^{(3)}(\mathbf{p}_1-\k)\delta^{(3)}(\k-\p_2)\delta_{s_1i}\rho_{js_2}(\k,t)
\nonumber \\ &+& 
(2\pi)^6 4k^0p_1^0\delta^{(3)}(\k-\p_2)\delta^{(3)}(\p_1-\k)
\nonumber \\ &\times & 
\delta_{s_2j}\rho_{s_1i}(\p_1,t_1,t)
~.
\eea




\appendix

\end{document}